\documentclass[12inch,a4paper]{article}
\usepackage[top=3cm, bottom=2.5cm, left=2.4cm, right=2.7cm]{geometry}
\usepackage{graphicx}
\usepackage{epstopdf}
\usepackage{booktabs}
\usepackage{blkarray}
\usepackage[british]{babel}

 \usepackage[sort,comma,authoryear]{natbib} 

    \setcitestyle{citesep={;}}
\setcitestyle{square}

\usepackage{amssymb}
\usepackage{setspace}
\usepackage{amsfonts}
\usepackage{amsmath}
\usepackage{float}
\usepackage{color}
\usepackage[dvipsnames]{xcolor}
 \usepackage{colortbl}
\usepackage{url}

\definecolor{Crimson}{rgb}{0.6471, 0.1098, 0.1882}

\usepackage{epstopdf}

\usepackage{tcolorbox}
\usepackage{epstopdf}

\usepackage{subcaption}
\providecommand{\keywords}[1]{\textit{Key words:} #1}

\newtheorem{definition}{\normalfont \color{Crimson} \textit{Definition}}[section]
\newtheorem{remark}{\normalfont \color{Crimson} \textit{Remark}}[section]
\newtheorem{theorem}{\color{Crimson} Theorem}[section]

\newtheorem{corollary}{\color{Crimson} Corollary}[section]
\newtheorem{lemma}{\color{Crimson} Lemma}[section]

\usepackage[font=small,labelfont=bf,
   justification=justified,
   format=plain]{caption}

\usepackage{titlesec}
\titleformat*{\subsection}{\normalfont \large}
\usepackage{abstract}
\usepackage{slashbox}

\begin{document}



\title{Classical and bayesian componentwise predictors for non-compact correlated ARH(1) processes}

\author{M. Dolores Ruiz--Medina$^1$ and Javier \'Alvarez-Li\'ebana$^1$}
\maketitle
\begin{flushleft}
$^1$ Department of Statistics and O. R., University of Granada, Spain.

\textit{E-mail: javialvaliebana@ugr.es}
\end{flushleft}

\doublespacing



\renewcommand{\absnamepos}{flushleft}
\setlength{\absleftindent}{0pt}
\setlength{\absrightindent}{0pt}
\renewcommand{\abstractname}{Summary}
\begin{abstract}
A special class of standard Gaussian Autoregressive Hilbertian processes of order one (Gaussian ARH(1) processes), with bounded linear autocorrelation operator, which does not satisfy the usual Hilbert-Schmidt assumption, is considered. To compensate the slow decay of the diagonal coefficients of the autocorrelation operator, a faster decay velocity of the eigenvalues of the trace autocovariance operator of the innovation process is assumed. As usual, the eigenvectors of the autocovariance operator of the ARH(1) process are considered for projection, since, here, they are assumed to be known. Diagonal componentwise classical and bayesian estimation of the autocorrelation operator is studied for prediction. The asymptotic efficiency and equivalence of both estimators is proved, as well as of their associated componentwise ARH(1) plugin predictors. A simulation study is undertaken to illustrate the theoretical results derived.

\vspace{0.5cm}
Published in \textbf{REVSTAT (in press)}

 \url{https://www.ine.pt/revstat/pdf/Classicalandbayesiancomponentwise.pdf}
\end{abstract}


\textcolor{Crimson}{\section{Introduction}
\label{A5:sec:1}}

Functional time series theory plays a key role in the  analysis of high-dimensional data (see, for example, \cite{Aueetal15,Bosq00,BosqBlanke07}). Inference for stochastic processes can also be addressed from this framework
(see \cite{Alvarezetal16} in relation to functional prediction of the Ornstein--Uhlenbeck process, in an ARH(1) process framework).  \cite{Bosq00} addresses the problem of
infinite--dimensional parameter estimation and prediction of
 ARH(1) processes, in the cases of known and unknown eigenvectors of the autocovariance operator.  Alternative
 projection methodologies have  been adopted, for example,  in \cite{AntoniadisSapatinas03}, in terms of  wavelet
bases,  and \cite{BesseCardot96}, in terms of spline bases.
The book by \cite{BosqBlanke07} provides a general overview on
statistical prediction, including Bayesian predictors, inference by projection and  kernel methods, empirical
density estimation, and linear processes in high--dimensional spaces (see also \cite{BlankeBosq15} on Bayesian prediction for stochastic
processes).
Recently, \cite{BosqRuiz14} have derived new results on
asymptotic efficiency and equivalence of classical and Bayes
predictors for $l^{2}$--valued Poisson process, where, as usual,
$l^{2}$ denotes the Hilbert space of square summable sequences.
Classical and Bayesian componentwise parameter
estimators of the mean function and autocovariance operator,
characterizing  Gaussian measures in Hilbert spaces, are also
compared in terms of their asymptotic efficiency, in that paper.

We first recall that the class of processes studied here could be of
interest in applications,  for instance, in the context of anomalous
physical diffusion processes (see, for example,  \cite{GorenfloMainardi03,Meerschaertetal02,MetzlerKlafter04}, and the references therein). An interesting example of our framework corresponds to the case of spatial fractal  diffusion operator, and regular innovations. Specifically, the class of  standard Gaussian ARH(1) processes studied
have a bounded linear autocorrelation operator, admitting  a weak--sense  diagonal spectral representation, in terms
 of the eigenvectors of the autocovariance operator. The sequence of diagonal coefficients, in such a spectral
representation, displays an accumulation point at one. The singularity of the autocorrelation kernel is compensated by the regularity of the autocovariance kernel of the innovation process.
Namely,  the key assumption here is the  summability of the  quotient between the eigenvalues of the autocovariance operator of the innovation process and of the  ARH(1) process. Under suitable conditions, the
 asymptotic efficiency and equivalence of the studied diagonal componentwise classical and Bayesian estimators of the
 autocorrelation operator are derived  (see \textcolor{Crimson}{Theorem} \ref{A5:mr} below). Under the same setting of conditions, the asymptotic efficiency and equivalence of the corresponding classical and Bayesian ARH(1) plug--in predictors are proved as well
 (see \textcolor{Crimson}{Theorem} \ref{A5:mr2} below). Although both theorems only refer to the case of known eigenvectors of the autocovariance operator, as illustrated in the simulation study undertaken in \cite{Alvarezetal17} (see also \cite{Alvarez17,RuizAlvarez17b}), a similar performance is obtained for the case of unknown eigenvectors, in comparison with
  other componentwise, kernel--based, wavelet-based penalized and nonparametric  approaches
adopted in  the current literature (see \cite{AntoniadisSapatinas03,BesseCardot96,Bosq00,Guillas01,Mas99}).

  Note that, for $\theta $ being
the unknown parameter, in order to compute ${\rm E} \left\lbrace \theta | X_{1},\ldots, X_{n} \right\rbrace,$ with \linebreak $\left\lbrace X_{1},\ldots, X_{n} \right\rbrace$ denoting the
functional sample, we suppose that $$\theta_{j} \bot \left\lbrace
X_{i,j^{\prime }},\ i\geq 1, j^{\prime }\neq j \right\rbrace,$$ which leads
to
$$\langle {\rm E} \left\lbrace \theta
|X_{1},\ldots,X_{n}\right\rbrace,v_{j}\rangle_{H}={\rm E} \left\lbrace \theta_{j}
|X_{1},\ldots,X_{n}\right\rbrace = {\rm E} \left\lbrace \theta_{j}
|X_{1,j},\ldots,X_{n,j}\right\rbrace.$$

Here, for each $j\geq 1,$ $\theta_{j}=\langle
\theta,v_{j} \rangle_{H},$ and $X_{i,j}=\langle
X_{i},v_{j}\rangle_{H},$ for each $i=1,\dots,n,$ with $\langle\cdot,\cdot \rangle_{H}$ being the
inner product in the real separable  Hilbert space $H$. Note that  $\left\lbrace v_{j},\ j\geq 1 \right\rbrace$ denotes  an orthonormal
basis   of $H,$ diagonalizing the common
autocovariance operator of $\left( X_{1},\ldots, X_{n} \right).$ We  can then
perform an independent computation of the respective posterior
distributions of the projections $\left\lbrace \theta_{j}, \ j\geq 1 \right\rbrace,$ of
parameter $\theta ,$ with respect to the orthonormal basis
$\left\lbrace v_{j},\ j\geq 1 \right\rbrace$ of $H.$

 Finally, some numerical examples are considered to illustrate the results derived on asymptotic efficiency and equivalence
 of moment--based  classical and Beta--prior--based Bayes diagonal componentwise  parameter estimators, and the associated ARH(1) plug--in predictors.

%
%
%

\textcolor{Crimson}{\section{Preliminaries}
\label{A5:preliminares}}

 The preliminary  definitions and results needed in the subsequent development  are introduced in this section.  We first refer to  the usual  class
of standard ARH(1) processes introduced in  \cite{Bosq00}.

\bigskip

\begin{definition} 
\label{A5:defsc}  
Let $H$ be a real separable
Hilbert space. A sequence $Y=\left\lbrace Y_{n},  \ n\in \mathbb{Z} \right\rbrace$ of
$H$--valued random variables on a basic probability space $(\Omega,
\mathcal{A},\mathcal{P})$ is called an autoregressive Hilbertian process of
order one, associated with $(\mu, \varepsilon, \rho),$ if it is
stationary and satisfies
\begin{equation}
\label{A5:modelarh0}
X_{n}=Y_{n}- \mu=\rho(Y_{n-1}-\mu)
+\varepsilon_{n}=\rho(X_{n-1})+\varepsilon_{n}, \quad n \in
\mathbb{Z},
\end{equation}
\noindent where $\varepsilon=\left\lbrace \varepsilon_{n}, \  n \in \mathbb{Z} \right\rbrace$ is a Hilbert--valued  white noise in the strong sense (i.e., a
 zero--mean  stationary sequence of independent $H-$valued random variables with ${\rm E} \left\lbrace \|
\varepsilon_{n}\|^{2}_{H} \right\rbrace=\sigma^{2}<\infty,$ for every $n\in \mathbb{Z}$),  and $\rho \in
\mathcal{L}(H),$ with $\mathcal{L}(H)$ being the space of linear
bounded operators on $H.$ For each $n\in \mathbb{Z},$ $\varepsilon_{n}$ and
$X_{n-1}$ are assumed to be uncorrelated.
\end{definition}

\bigskip

If there exists a positive $j_{0}\geq 1$ such that
$\|\rho^{j_{0}}\|_{\mathcal{L}(H)}<1,$ then, the ARH(1) process in (\ref{A5:modelarh0}) is standard, and  there exists  a unique
stationary solution to equation (\ref{A5:modelarh0}) admitting a MAH($\infty $) representation (see \cite[Theorem 3.1, p. 74]{Bosq00}).

The autocovariance and cross--covariance operators are given, for each $n \in \mathbb{Z}$, by
\begin{eqnarray}
C &=& {\rm E} \left\lbrace X_{n}\otimes X_{n} \right\rbrace = {\rm E} \left\lbrace X_{0}\otimes X_{0} \right\rbrace, \quad D=  {\rm E} \left\lbrace X_{n}\otimes X_{n+1} \right\rbrace = {\rm E} \left\lbrace X_{0} \otimes X_{1} \right\rbrace,
\label{A5:opcov}
\end{eqnarray}
\noindent where, for $f,g\in H,$
$$f\otimes g(h)=f\left\langle g,h\right\rangle_{H},\quad \forall h\in H,$$ defines a Hilbert--Schmidt operator on $H.$ The operator $C$ is
assumed to be in the trace class. In particular,
$${\rm E} \left\lbrace \|X_{n}\|^{2}_{H} \right\rbrace <\infty, \quad n\in \mathbb{Z}.$$ It is well-known that, from equations (\ref{A5:modelarh0})--(\ref{A5:opcov}), for all $h\in H,$ $D(h)= \rho C(h)$ (see, for example, \cite{Bosq00}).
However, since $C$ is a nuclear or trace operator,  its inverse
operator is an unbounded operator in $H.$ Different methodologies
have been adopted to overcome this problem in the current literature
on ARH(1) processes. In particular, here, we consider the case where $C(H)=H,$ under  \textcolor{Aquamarine}{\textbf{Assumption A2}} below,  since
$C$ is assumed to be strictly positive. That is, its   eigenvalues are strictily positive and the kernel space of $C$ is trivial.  In addition, they are assumed to have   multiplicity one. Therefore, for any $f,g\in H,$ there exist $\varphi, \phi  \in H$ such that $f =C(\varphi )$ and $g= C(\phi),$ and $$
\left\langle C^{-1}(f),C^{-1}(g)\right\rangle_{H}=\left\langle C^{-1}(C(\varphi )),C^{-1}(C(\phi))\right\rangle_{H}=
\left\langle \varphi, \phi\right\rangle_{H}.$$

In particular, $$\|C^{-1}(f)\|_{H}^{2}<\infty , \quad \forall f\in H.$$

\bigskip

\noindent \textcolor{Aquamarine}{\textbf{Assumption A1.}} The operator $\rho $ in
(\ref{A5:modelarh0}) is  self--adjoint with
$\|\rho\|_{\mathcal{L}(H)}<1.$

\bigskip

\noindent \textcolor{Aquamarine}{\textbf{Assumption  A2.}} The operator $C$ is
strictly positive, and its positive eigenvalues have multiplicity one. Furthermore, $C$ and $\rho$ admit the
following  diagonal spectral decompositions, such that for all $f,g\in H,$
\begin{eqnarray}
C(g)(f)&=& \displaystyle \sum_{k=1}^{\infty}C_k \left\langle\phi_{k},g\right\rangle_{H}\left\langle\phi_{k},f\right\rangle_{H}\label{A5:eqrecovttt}\\
\rho (g)(f)&=& \displaystyle \sum_{k=1}^{\infty}\rho_{k} \left\langle\phi_{k},
g\right\rangle_{H}\left\langle\phi_{k},f\right\rangle_{H},
\label{A5:eqrerhott}
\end{eqnarray}
\noindent where $\{C_k,\ k\geq
1\}$ and $\{\rho_{k},\ k\geq 1\}$ are the respective systems of eigenvalues of $C$ and $\rho,$ and \linebreak $\{\phi_{k},\
k\geq 1\}$ is the common system of orthonormal eigenvectors of the autocovariance operator $C.$

\bigskip

\begin{remark}
\textit{As commented before, we consider here the case where the eigenvectors $\left\lbrace \phi_{k}, \ k\geq 1 \right\rbrace$ of the autocovariance operator $C$ are known. Thus, under  \textcolor{Aquamarine}{\textbf{Assumption A2}}, the natural way to formulate a  componentwise estimator of the autocorrelation operator  $\rho$ is in terms of the respective estimators of its diagonal coefficients $\left\lbrace \rho_{k}, \ k\geq 1 \right\rbrace,$ computed from the respective projections of the observed functional data, $\left( X_{0},\ldots,X_{T} \right),$ into $\{\phi_{k},\
k\geq 1\}$. We adopt here a moment--based  classical  and Beta--prior--based Bayesian approach in the
estimation of such  coefficients $\left\lbrace \rho_{k}, \ k\geq 1 \right\rbrace.$}
\end{remark}

\bigskip

From the Cauchy--Schwarz's inequality, applying the Parseval's identity, 
\begin{eqnarray}
\left|\rho (g)(f)\right|^{2}&\leq
&\displaystyle \sum_{k=1}^{\infty}|\rho_{k}|\left[\left\langle\phi_{k},
g\right\rangle_{H}\right]^{2} \displaystyle \sum_{k=1}^{\infty}|\rho_{k}|\left[\left\langle\phi_{k}, f\right\rangle_{H}\right]^{2}\nonumber\\
&\leq & \displaystyle \sum_{k=1}^{\infty}\left[\left\langle\phi_{k},
g\right\rangle_{H}\right]^{2} \displaystyle \sum_{k=1}^{\infty}\left[\left\langle\phi_{k},
f\right\rangle_{H}\right]^{2}=\|g\|^{2}_{H}\|f\|^{2}_{H}<\infty.\nonumber
\end{eqnarray}

Thus, equation (\ref{A5:eqrerhott}) holds in the weak sense.

\bigskip

   From \textcolor{Aquamarine}{\textbf{Assumption A2}}, the projection of $X_{n}$ into the common
eigenvector system $\{\phi_{k},\ k\geq 1\}$ leads to the following
series expansion in $\mathcal{L}^{2}_{H}(\Omega,\mathcal{A},\mathcal{P}):$
\begin{equation}
X_{n}=\displaystyle \sum_{k=1}^{\infty
}\sqrt{C_k}\eta_{k}(n)\phi_{k},\quad
\eta_{k}(n)=\frac{1}{\sqrt{C_k}}\left\langle
X_{n},\phi_{k}\right\rangle_{H},\label{A5:kesxp}
\end{equation}
\noindent and, for each $j,p\geq 1,$ and $n>0,$
\begin{eqnarray}
{\rm E} \left\lbrace \eta_{j}(n)\eta_{p}(n) \right\rbrace &=&
{\rm E} \left\lbrace \frac{1}{\sqrt{C_j}}\left\langle
X_{n},\phi_{j}\right\rangle_{H}\frac{1}{\sqrt{C_p}}
\left\langle X_{n},\phi_{p}\right\rangle_{H}\right\rbrace \nonumber\\
&=&\frac{1}{\sqrt{C_j}}\frac{1}{\sqrt{C_p}}C(\phi_{j})(\phi_{p})\nonumber\\&=&\frac{1}{\sqrt{C_j}}
\frac{1}{\sqrt{C_p}}
C_j \left\langle\phi_{j},\phi_{p}\right\rangle_{H}=\delta_{j,p}, \nonumber
\end{eqnarray}
\noindent where the last equality is obtained from the
orthonormality of the eigenvectors $\{\phi_{k},\ k\geq 1\}.$  Hence, under \textcolor{Aquamarine}{\textbf{Assumptions A1--A2}}, the projection
of equation (\ref{A5:modelarh0}) into the elements of the common
eigenvector system $\{\phi_{k},\ k\geq 1\}$ leads to the following
infinite-dimensional system of equations:

\begin{equation}
\sqrt{C_k}\eta_{k}(n)=\rho_{k}\sqrt{C_k}\eta_{k}(n-1)+\varepsilon_{k}(n),\quad
k\geq 1,\label{A5:eqardprev}
\end{equation}
\noindent or equivalently,

\begin{equation}
\eta_{k}(n)=\rho_{k}\eta_{k}(n-1)+\frac{\varepsilon_{k}(n)}{\sqrt{C_k}},\quad
k\geq 1,\label{A5:eqard}
\end{equation}
\noindent where $$\varepsilon_{k}(n)=\left\langle
\varepsilon_{n},\phi_{k}\right\rangle_{H}, \quad k\geq 1, \quad n\in
\mathbb{Z}.$$

 Thus,   for each $j\geq 1,$ $$\{a_{j}(n)=
\sqrt{C_j}\eta_{j}(n),\ n \in \mathbb{Z}\}$$ defines a
standard AR(1) process. Its  moving average representation of
infinite order  is given by
\begin{equation}
a_{j}(n)= \displaystyle \sum_{k=0}^{\infty
}[\rho_{j}]^{k}\varepsilon_{j}(n-k),\quad n\in \mathbb{Z}.
\label{A5:streq}
\end{equation}

Specifically, under \textcolor{Aquamarine}{\textbf{Assumption A2}},
 \begin{eqnarray}
 {\rm E} \left\lbrace a_{j}(n)a_{p}(n) \right\rbrace &=& \displaystyle \sum_{k=0}^{\infty
}\displaystyle \sum_{l=0}^{\infty
}[\rho_{j}]^{k}[\rho_{p}]^{l}{\rm E} \left\lbrace \varepsilon_{j}(n-k)\varepsilon_{p}(n-l) \right\rbrace
\nonumber\\
&=& \displaystyle \sum_{k=0}^{\infty } \displaystyle \sum_{l=0}^{\infty
}[\rho_{j}]^{k}[\rho_{p}]^{l}\delta_{k,l}\delta_{j,p}=0,\quad j\neq
p,\nonumber\\
{\rm E} \left\lbrace a_{j}(n)a_{p}(n) \right\rbrace &=&\displaystyle \sum_{k=0}^{\infty
}\sigma_{j}^{2}[\rho_{j}]^{2k},\quad j=p,\label{A5:variance}
\end{eqnarray}
\noindent where
$$\sigma_{j}^{2}={\rm E} \left\lbrace \varepsilon_{j}(n-k) \right\rbrace^{2}= {\rm E} \left\lbrace \varepsilon_{j}(0) \right\rbrace^{2}.$$

 From equation (\ref{A5:variance}), under \textcolor{Aquamarine}{\textbf{Assumptions A1--A2}},
\begin{eqnarray}
{\rm E} \left\lbrace \|X(n)\|^{2}_{H} \right\rbrace &=& \displaystyle\sum_{j=1}^{\infty} {\rm E} \left\lbrace a_{j}(n) \right\rbrace^{2}= \displaystyle \sum_{j=1}^{\infty
}\sigma_{j}^{2} \displaystyle \sum_{k=0}^{\infty }[\rho_{j}]^{2k}
\nonumber\\
&=& \displaystyle \sum_{j=1}^{\infty
}\sigma_{j}^{2}\left[\frac{1}{1-[\rho_{j}]^{2}}\right]= \displaystyle \sum_{j=1}^{\infty}C_j <\infty,\label{A5:varfunct}
\end{eqnarray}
\noindent
 with, as before,
$$\displaystyle \sum_{j=1}^{\infty }\sigma_{j}^{2}= {\rm E} \left\lbrace \|\varepsilon_n
\|^{2}_{H} \right\rbrace <\infty.
$$

Equation (\ref{A5:varfunct})  leads to the identity
\begin{equation}
C_j =\left[\frac{\sigma_{j}^{2}}{1-\rho^{2}_{j}}\right],\quad j\geq 1,\label{A5:id2rem}
\end{equation}
\noindent from which, we obtain

\begin{equation}
\rho_{k}=\sqrt{1-\frac{\sigma_{k}^{2}}{\lambda_{k}(C)}},\quad
\sigma_{k}^{2}= {\rm E} \left\lbrace \left\langle
\phi_{k},\varepsilon_{n}\right\rangle_{H} \right\rbrace^{2},\quad \forall
n\in \mathbb{Z}, \quad k\geq 1.\label{A5:eqreeigcrho}
\end{equation}

Under (\ref{A5:id2rem}), equation (\ref{A5:eqard}) can also be rewritten as

\begin{equation}\eta_{k}(n)=\rho_{k}\eta_{k}(n-1)+\sqrt{1-\rho_{k}^{2}}\frac{\varepsilon_{k}(n)}{\sigma_{k}},\quad
k\geq 1, \nonumber 
\end{equation}

\bigskip

\noindent  \textcolor{Aquamarine}{\textbf{Assumption A2B.}} The  sequences $$\left\lbrace \sigma^{2}_{k}, \ k\geq 1 \right\rbrace, \quad \left\lbrace C_k, \ k\geq 1 \right\rbrace$$
satisfy
\begin{eqnarray}
& & \frac{\sigma_{k}^{2}}{C_k}\leq 1,\ k\geq 1,\quad
\lim_{k\rightarrow \infty}\frac{\sigma_{k}^{2}}{C_k}=0, \nonumber\\
& & \frac{\sigma_{k}^{2}}{C_k}=\mathcal{O}(k^{-1-\gamma}),\quad
\gamma>0,\quad  k\rightarrow \infty. \nonumber \\ \label{A5:orderconvzeroinnv2}
\end{eqnarray}

Equation  (\ref{A5:orderconvzeroinnv2}) means that $\left\lbrace \sigma^{2}_{k}, \ k\geq 1 \right\rbrace$ and $ \left\lbrace C_k, \ k\geq 1 \right\rbrace$ are
both summable
 sequences, with faster decay to zero of the sequence $\left\lbrace \sigma^{2}_{k}, \ k\geq 1 \right\rbrace$ than the sequence $\left\lbrace C_k, \ k\geq 1 \right\rbrace,$ leading, from equations (\ref{A5:id2rem})--(\ref{A5:eqreeigcrho}),
 to the definition of $\left\lbrace \rho_{k}^{2}, \ k\geq 1 \right\rbrace$ as a sequence with accumulation point at one.

\bigskip

\begin{remark}
\label{A5:remcondA3}
\textit{Under \textcolor{Aquamarine}{\textbf{Assumption A2B}} and \textcolor{Aquamarine}{\textbf{A3}} below holds.}
\end{remark}

\bigskip

For each $k\geq 1,$ from  equations (\ref{A5:eqardprev})--(\ref{A5:streq}),
\begin{eqnarray}
\displaystyle \sum_{n=1}^{T}[\eta_{k}(n-1)]^{2} &=& \frac{1}{C_k}\left[\displaystyle \sum_{n=1}^{T}[\varepsilon_{k}(n-1)]^{2}\right.
\nonumber\\ &&\left.+ \displaystyle \sum_{n=1}^{T}\sum_{l=1}^{\infty }
\displaystyle \sum_{p=1}^{\infty
}[\rho_{k}]^{l}[\rho_{k}]^{p}\varepsilon_{k}(n-1-l)\varepsilon_{k}(n-1-p)\right]
\nonumber\\
&=&
\frac{1}{C_k}\left[\displaystyle \sum_{n=1}^{T}[\varepsilon_{k}(n-1)]^{2}+S(T,k)\right], \nonumber 
\end{eqnarray}
\noindent where $$S(T,k)= \displaystyle \sum_{n=1}^{T} \displaystyle \sum_{l=1}^{\infty }
\displaystyle \sum_{p=1}^{\infty }[\rho_{k}]^{l}[\rho_{k}
]^{p}\varepsilon_{k}(n-1-l)\varepsilon_{k}(n-1-p).$$ Hence,
$\displaystyle \sum_{n=1}^{T}[\varepsilon_{k}(n-1)]^{2}+S(T,k)\geq 0,$ for every
$T\geq 1,$ and $k\geq 1.$

\bigskip

\noindent \textcolor{Aquamarine}{\textbf{Assumption A3.}} There
exists a sequence of real-valued independent random variables   $\left\lbrace \widetilde{M}(k), \ k\geq 1 \right\rbrace$ such that
\begin{eqnarray}
\displaystyle \inf_{T\geq 1}\sqrt{\left|\frac{S(T,k)}{T\left(\displaystyle \sum_{n=1}^{T-1}\left[\varepsilon_{k}
(n)\right]^{2}+\left[\varepsilon_{k} (0)\right]^{2}\right)}\right|}
 &=& \displaystyle \inf_{T\geq
1}\sqrt{\left|\frac{\displaystyle \sum_{n=1}^{T} \displaystyle \sum_{l=1}^{\infty } \displaystyle \sum_{p=1}^{\infty
}[\rho_{k}]^{l}[\rho_{k}]^{p}\varepsilon_{k}(n-1-l)\varepsilon_{k}(n-1-p)}{T\left( \displaystyle \sum_{n=1}^{T-1}\left[\varepsilon_{k}
(n)\right]^{2}+\left[\varepsilon_{k} (0)\right]^{2}\right)}\right|} \nonumber\\
& \geq & [\widetilde{M}(k)]^{-1}~a.s., \nonumber
\end{eqnarray}
\noindent with
\begin{eqnarray}
\displaystyle \sum_{k=1}^{\infty}{\rm E} \left\lbrace \widetilde{M}(k) \right\rbrace^{l}<\infty,\quad 1\leq l\leq 4.\label{A5:asi22}
\end{eqnarray}

\bigskip

\begin{remark}
\label{A5:remark_new_}
\textit{Note that the mean value of $$\displaystyle \sum_{n=1}^{T}\displaystyle \sum_{l=1}^{\infty }
\displaystyle \sum_{p=1}^{\infty
}[\rho_{k}]^{l}[\rho_{k}]^{p}\varepsilon_{k}(n-1-l)\varepsilon_{k}(n-1-p)$$ is of order $\frac{T\sigma_{k}^{2}}{1-(\rho_{k})^{2}},$
and the mean value of
$$T\left(\sum_{n=1}^{T-1}\left[\varepsilon_{k}
(n)\right]^{2}+\left[\varepsilon_{k} (0)\right]^{2}\right)$$
  is of order $T(T-1)\sigma_{k}^{2}.$ Hence, for the almost
surely boundedness of the inverse of
$$\left|\frac{S(T,k)}{T\left(\displaystyle \sum_{n=1}^{T-1}\left[\varepsilon_{k}
(n)\right]^{2}+\left[\varepsilon_{k} (0)\right]^{2}\right)}\right|,$$
by a suitable sequence of random variables with summable
$l$--moments, for $l=1,2,3,4,$ the eigenvalues of operator $\rho $
must be  close to one but strictly less than one. As commented in \textcolor{Crimson}{Remark} \ref{A5:remcondA3}, from \textcolor{Aquamarine}{\textbf{Assumption A2B}}, this condition
 is satisfied in view of equation (\ref{A5:eqreeigcrho}).}
\end{remark}

\bigskip

\noindent \textcolor{Aquamarine}{\textbf{Assumption A4.}}
${\rm E} \left\lbrace \eta_{j}(m)\eta_{k}(n) \right\rbrace =\delta_{j,k},$ with, as before,
$\delta_{j,k}$ denoting the Kronecker delta  function, for every
$m,n\in \mathbb{Z},$ and $j,k\geq 1.$

\bigskip

\begin{remark}
\textit{\textcolor{Aquamarine}{\textbf{Assumption A4}} implies that the cross--covariance operator $D$ admits a diagonal spectral decomposition in terms of the system of eigenvectors $\left\lbrace \phi_{k}, \ k\geq 1 \right\rbrace.$ Thus, under \textcolor{Aquamarine}{\textbf{Assumption A4}}, the diagonal spectral decompositions (\ref{A5:eqrecovttt})--(\ref{A5:eqrerhott}) also hold.}
\end{remark}

\bigskip

The classical diagonal componentwise  estimator $\widehat{\rho}_{T}$ of
$\rho$ considered here is given by
\begin{eqnarray}
\widehat{\rho}_{T}&=&\displaystyle \sum_{k=1}^{\infty }\widehat{\rho}_{k,T}\phi_{k}\otimes \phi_{k}\nonumber\\
\widehat{\rho}_{k,T}&=&
\frac{\displaystyle \sum_{n=1}^{T}a_{k}(n-1)a_{k}(n)}{\displaystyle \sum_{n=1}^{T}[a_{k}(n-1)]^{2}} =  \frac{\displaystyle \sum_{n=1}^{T}\left\langle
X_{n-1},\phi_{k}\right\rangle_{H}\left\langle
X_{n},\phi_{k}\right\rangle_{H}}{\displaystyle \sum_{n=1}^{T}\left[\left\langle
X_{n-1},\phi_{k}\right\rangle_{H}\right]^{2}}\nonumber\\
&=&\frac{\displaystyle \sum_{n=1}^{T} X_{n-1,k} X_{n,k}}{\displaystyle \sum_{n=1}^{T}
X^{2}_{n-1,k}},\quad k\geq 1.
 \label{A5:componentwisee}
\end{eqnarray}

From equations (\ref{A5:eqardprev})--(\ref{A5:eqard}) and (\ref{A5:id2rem}), for each  $k\geq 1,$
\begin{eqnarray}
\widehat{\rho}_{k,T}-\rho_{k}&=&\frac{\displaystyle \sum_{n=1}^{T}X_{n-1,k}X_{n,k}}{\displaystyle \sum_{n=1}^{T}[X_{n-1,k}]^{2}}-\rho_{k}\nonumber\\
&=&\frac{\displaystyle \sum_{n=1}^{T}\rho_{k}[\eta_{k}(n-1)]^{2}+(\eta_{k}(n-1)\varepsilon_{k}(n))/\sqrt{C_k}}{\displaystyle \sum_{n=1}^{T}[\eta_{k}(n-1)]^{2}}
-\rho_{k}\nonumber \\
&=&\rho_{k}+\frac{\displaystyle \sum_{n=1}^{T}\eta_{k}(n-1)\varepsilon_{k}(n)}{\sqrt{C_k}\displaystyle \sum_{n=1}^{T}[\eta_{k}(n-1)]^{2}}-\rho_{k}\nonumber
\end{eqnarray}
\begin{eqnarray}
&=&\frac{\displaystyle \sum_{n=1}^{T}\eta_{k}(n-1)\varepsilon_{k}(n)}{\sqrt{\sigma_{k}^{2}/(1-\rho_{k}^{2})}\displaystyle \sum_{n=1}^{T}[\eta_{k}(n-1)]^{2}} \nonumber \\ 
&=&
\sqrt{1-\rho_{k}^{2}}\frac{\displaystyle \sum_{n=1}^{T}\eta_{k}(n-1)[\varepsilon_{k}(n)/\sigma_{k}]}{\displaystyle \sum_{n=1}^{T}[\eta_{k}(n-1)]^{2}}.
 \label{A5:eqest}
\end{eqnarray}

\bigskip

\begin{remark}
\textit{
It is important
to note that, for instance, unconditional bases, like wavelets, provide
the spectral diagonalization of an extensive family of operators, including
pseudodifferential operators, and in particular, Calder\'on--Zygmund
operators (see \cite{KyriazisPetrushev01,MeyerCoifman97}). Therefore, the diagonal spectral representations (\ref{A5:eqrecovttt})--(\ref{A5:eqrerhott}), in \textcolor{Aquamarine}{\textbf{Assumption A2}}, hold for a wide class of autocovariance
and cross-covariance operators, for example, in terms of wavelets.
When the autocovariance and the cross--covariance operators are related
by a continuous function, the diagonal spectral representations (\ref{A5:eqrecovttt})--(\ref{A5:eqrerhott}) are also satisfied (see \cite[pp. 119, 126 and 140]{DautrayLions90}). \textcolor{Aquamarine}{\textbf{Assumption A2}} has  been considered, for example, in \cite[Theorem 8.5, pp. 215--216; Theorem 8.7, p. 221]{Bosq00}, to establish strong consistency, although, in this book, a different setting of conditions is assumed. Thus, \textcolor{Aquamarine}{\textbf{Assumptions A1--A2}} already have been used (e.g., in \cite{Bosq00,Alvarezetal17,RuizAlvarez17b}),  and \textcolor{Aquamarine}{\textbf{Assumptions A2B, A3}} and \textcolor{Aquamarine}{\textbf{A4}}   appear in \cite{Ruizetal16}.  \textcolor{Aquamarine}{\textbf{Assumptions A2B}} is needed since  the usual assumption on the Hilbert--Schmidt property of $\rho,$ made by several authors, is not considered here. At the same type, as commented before, \textcolor{Aquamarine}{\textbf{Assumptions A2B}}  implies \textcolor{Aquamarine}{\textbf{Assumption A3}}}.
\end{remark}

\bigskip

The following lemmas will be used in the derivation of the main results of this paper, \textcolor{Crimson}{Theorems} \ref{A5:mr} and \ref{A5:mr2}, obtained in the Gaussian ARH(1) context.

\bigskip

\begin{lemma}
\label{A5:prdistab} 
\textit{Let $\left\lbrace \mathcal{X}_{i}, \ i=1,\dots,n \right\rbrace,$ be the values
of a  standard zero--mean autoregressive process of order one (AR(1)
process) at times $i=1,\dots,n,$ and
$$\widehat{\rho}_{n}=\frac{\displaystyle \sum_{i=1}^{n}\mathcal{X}_{i-1}\mathcal{X}_{i}}{\displaystyle \sum_{i=1}^{n}\mathcal{X}_{i-1}^{2}},$$
with $\mathcal{X}_{1}$ representing the random initial condition. Assume that $|\rho|<1,$ and that the innovation process is white noise.
Then, as $n\rightarrow \infty ,$
\begin{equation}\sqrt{n}\frac{\widehat{\rho}_{n}-\rho}{\sqrt{1-\rho^{2}}}\underset{\mathcal{L}}{\longrightarrow}\mathcal{N}(0,1). \nonumber 
\end{equation}}
\end{lemma}

\bigskip

The proof of \textcolor{Crimson}{Lemma} \ref{A5:prdistab} can be found in \cite[p. 216]{Hamilton94}.

\bigskip

\begin{lemma}
\label{A5:lem1}  
\textit{Let $\mathcal{X}_{1}$ and $\mathcal{X}_{2}$ be two
normal distributed random variables having correlation
$\rho_{\mathcal{X}_{1}\mathcal{X}_{2}},$ and with means $\mu_{1}$
and $\mu_{2},$  and variances $\sigma_{1}^{2}$ and $\sigma_{2}^{2},$
respectively. Then, the following identities hold:
\begin{eqnarray}
{\rm E} \left\lbrace \mathcal{X}_{1}\mathcal{X}_{2} \right\rbrace &=& \mu_{1}\mu_{2}+\rho_{\mathcal{X}_{1}\mathcal{X}_{2}}\sigma_{1}\sigma_{2}\nonumber\\
{\rm Var} \left\lbrace \mathcal{X}_{1}\mathcal{X}_{2} \right\rbrace &=&\mu_{1}^{2}\sigma_{2}^{2}+\mu_{2}^{2}\sigma_{1}^{2}+\sigma_{1}^{2}\sigma_{2}^{2}
+2\rho_{\mathcal{X}_{1}\mathcal{X}_{2}}
\mu_{1}\mu_{2}\sigma_{1}\sigma_{2}+\rho^{2}_{\mathcal{X}_{1}\mathcal{X}_{2}}\sigma_{1}^{2}\sigma_{2}^{2}
\label{A5:eqmomentnrvs}
\end{eqnarray}
\noindent (see, for example,  \cite{Aroian47,WareLad03}).}
\end{lemma}

\bigskip

\begin{lemma}
\label{A5:prdistab_2} 
\textit{For each $k\geq 1,$ the following limit is obtained:
\begin{equation}
\displaystyle \lim_{T\rightarrow \infty
}T {\rm E} \left\lbrace \widehat{\rho}_{k,T}-\rho_{k} \right\rbrace^{2}=1-\rho_{k}^{2},\quad k\geq
1 \label{A5:Bartlettf}
\end{equation} 
\noindent
(see, for example, \cite{Bartlett46}).}
\end{lemma}

\textcolor{Crimson}{\section{Bayesian diagonal componentwise estimation}\label{A5:ldc}}

Now let us denote by $R$ the functional random variable on the basic
probability space $(\Omega ,\mathcal{A},\mathcal{P}),$  characterized by the
prior distribution  for $\rho .$  In our case, we assume that $R$ is
of the form
\begin{equation}
R(f)(g)= \displaystyle \sum_{k=1}^{\infty} R_{k}\left\langle \phi_{k},f\right\rangle_{H}\left\langle \phi_{k},g\right\rangle_{H}~a.s.,\quad \forall f,g \in H, \nonumber 
\end{equation}
\noindent where, for $k\geq 1,$ $R_{k}$ is a real--valued random variable such that    $R(\phi_{j})(\phi_{k})=\delta_{j,k}R_{k},$ almost surely, for every $j\geq 1.$  In the following,  $R_{k}$ is  assumed to follow a   beta distribution with shape parameters
$a_{k}>0$ and $b_{k}>0$; i.e., $R_{k} \sim \mathcal{B}(a_{k},b_{k}),$ for every
$k\geq 1.$  We also assume that $R$ is independent of the functional components of the innovation process $\left\lbrace \varepsilon_{n},\ n\in \mathbb{Z} \right\rbrace,$
and that the random variables $\left\lbrace R_{k}, \ k\geq 1 \right\rbrace,$ are globally independent. That is, for each $f,g\in H,$
\begin{eqnarray}
\varphi_{R}^{f,g}(t)&=& {\rm E} \left\lbrace \exp\left( it\sum_{k=1}^{\infty} R_{k}\left\langle \phi_{k},f\right\rangle_{H}\left\langle \phi_{k},g\right\rangle_{H}\right)\right\rbrace\nonumber\\
&=& \displaystyle \prod_{k=1}^{\infty} {\rm E} \left\lbrace \exp\left( it R_{k}\left\langle \phi_{k},f\right\rangle_{H}\left\langle \phi_{k},g\right\rangle_{H}\right)\right\rbrace
= \displaystyle \prod_{k=1}^{\infty}\varphi_{R_{k}}\left(t\left\langle \phi_{k},f\right\rangle_{H}\left\langle \phi_{k},g\right\rangle_{H}\right).
\label{A5:eqcharactfunct}
\end{eqnarray}

Thus,
$$\varphi_{R}(t)=\displaystyle \prod_{k=1}^{\infty}\varphi_{R_{k}}\left(t\left( \phi_{k}\otimes \phi_{k}\right)\right),$$
\noindent where the last identity is understood in the weak--sense; i.e., in the sense of  equation (\ref{A5:eqcharactfunct}).
 
 In the definition of $R$ from $\{R_{j},\ j\geq 1\},$ we  can then
apply the Kolmogorov extension Theorem under the condition
$$\displaystyle \sum_{j=1}^{\infty
}\frac{a_{j}b_{j}}{(a_{j}+b_{j}+1)(a_{j}+b_{j})^{2}}<\infty$$\noindent  (see, for example, \cite{Khoshnevisan07}).

As in the real--valued  case (see \textcolor{Crimson}{Supplementary Material} \ref{A5:Supp1}),  considering
$b_{j}>1,$ for each $j\geq 1,$   the Bayes estimator of $\rho $ is defined by (see Case 2 in \textcolor{Crimson}{Supplementary Material} \ref{A5:Supp1})
\begin{equation}\widetilde{\rho }_{T}=\sum_{j=1}^{\infty }\widetilde{\rho
}_{j,T}\phi_{j}\otimes \phi_{j},\label{A5:Berho}
\end{equation}

\noindent with, for every $j\geq 1,$
\begin{eqnarray}
\widetilde{\rho }_{j,T}&=&\frac{1}{2\beta_{j,T} }\left[
(\alpha_{j,T} +\beta_{j,T} )\pm \sqrt{(\alpha_{j,T} -\beta_{j,T}
)^{2}-4\beta_{j,T}
\sigma_{j}^{2}[2-(a_{j}+b_{j})]}\right]\nonumber\\
&=&
\frac{\left[\displaystyle \sum_{i=1}^{T}x_{i-1,j}x_{i,j}+x_{i-1,j}^{2}\right]}{2\displaystyle \sum_{i=1}^{T}x_{i-1,j}^{2}}\nonumber\\
&\pm &
\frac{\sqrt{\left[\displaystyle \sum_{i=1}^{T}x_{i-1,j}x_{i,j}-x_{i-1,j}^{2}\right]^{2}-4\sigma_{j}^{2}\left[\displaystyle \sum_{i=1}^{T}x_{i-1,j}^{2}\right]
[2-(a_{j}+b_{j})]}}{2\displaystyle \sum_{i=1}^{T}x_{i-1,j}^{2}},
\label{A5:Bayestrhobeta}
\end{eqnarray}

\noindent where 
\begin{equation}
\alpha_{j,T}=\sum_{i=1}^{T}x_{i-1,j}x_{i,j}, \quad
\beta_{j,T}=\sum_{i=1}^{T}x_{i-1,j}^{2},\ j\geq 1,\ n\geq 2.\label{A5:eqcoefbayest}
\end{equation}

\textcolor{Crimson}{\section{Asymptotic efficiency and equivalence} \label{A5:new_section}}

 In this section, sufficient conditions are derived to ensure
 the asymptotic efficiency and equivalence of the diagonal componentwise estimators of $\rho$ formulated in the classical
(see equation (\ref{A5:componentwisee})), and in the Bayesian  (see equations (\ref{A5:Berho})--(\ref{A5:eqcoefbayest})) frameworks.

\bigskip

\begin{theorem}
\label{A5:mr} 
\textit{Under  \textcolor{Aquamarine}{\textbf{Assumptions A1--A2, A2B, A3}} and \textcolor{Aquamarine}{\textbf{A4}}, let us assume that the ARH(1) process $X$ satisfies, for each $j\geq 1,$ and,  for every $T\geq 2,$
\begin{eqnarray}
\displaystyle \sum_{i=1}^{T}\varepsilon_{j}(i)X_{i-1,j}\geq 0,~a.s.\label{A5:eq2th1}
\end{eqnarray}
\noindent That is, $\{\varepsilon_{j}(i),\ i\geq 1\} $ and $\{X_{i-1,j}, \ i\geq 0\}$ are almost surely  positive empirically
 correlated. In addition,  for every  $j\geq 1,$ the hyper--parameters $a_{j}$ and $b_{j}$ of the beta prior
distribution, $\mathcal{B} (a_{j},b_{j}),$ are  such that $a_{j}+b_{j}\geq 2.$ Then, the following identities are obtained:
\begin{eqnarray}
\lim_{T\rightarrow \infty}
T {\rm E} \left\lbrace \|\widetilde{\rho}_{T}^{-}-\rho\|^{2}_{\mathcal{S}(H)}\right\rbrace = \displaystyle \lim_{T\rightarrow
\infty}
T{\rm E} \left\lbrace \|\widehat{\rho}_{T}-\rho\|^{2}_{\mathcal{S}(H)}\right \rbrace = \displaystyle \sum_{k=1}^{\infty
}\frac{\sigma_{k}^{2}}{C_k}<\infty,
\label{A5:eqfeqtheorem1}
\end{eqnarray}
\noindent where $\widehat{\rho}_{T}$ is defined in equation
(\ref{A5:componentwisee}), and $\widetilde{\rho}_{T}^{-}$ is defined from equations
(\ref{A5:Berho})--(\ref{A5:eqcoefbayest}), considering
\begin{equation}
\widetilde{\rho}_{j,T}^{-}=\frac{1}{2\beta_{j,T} }\left[
(\alpha_{j,T} +\beta_{j,T} )-\sqrt{(\alpha_{j,T} -\beta_{j,T}
)^{2}-4\beta_{j,T} \sigma_{j}^{2}[2-(a_{j}+b_{j})]}\right],\label{A5:eqestbayes}
\end{equation}
 \noindent with, as before, for each $j\geq 1,$ $$X_{i,j}=\left\langle
 X_{i},\phi_{j}\right\rangle_{H}, \quad i=0,\dots,T,$$  and $\alpha_{j,T}$ and  $\beta_{j,T}$ are given in (\ref{A5:eqcoefbayest}), for every $T\geq 2.$}
\end{theorem}

\begin{proof}
Under \textcolor{Aquamarine}{\textbf{Assumptions A1--A2}}, from \textcolor{Crimson}{Remark} \ref{A5:remark2} and \textcolor{Crimson}{Corollary} \ref{A5:corollary1} in  \textcolor{Crimson}{Supplementary Material} \ref{A5:Supp2}, for each $j\geq 1,$ and for $T$ sufficiently large,
\begin{equation}
|\widehat{\rho}_{j,T}|\leq 1,\quad \mbox{a.s.} \nonumber 
\end{equation}

Also, under (\ref{A5:eq2th1}),
$$\displaystyle \sum_{i=1}^{T}\rho_{j}X_{i-1,j}^{2}+\varepsilon_{j}(i)X_{i-1}\geq
\displaystyle \sum_{i=1}^{T}\rho_{j}X_{i-1,j}^{2},\quad \mbox{a.s.},$$ \noindent
which is equivalent to
\begin{equation}
\widehat{\rho}_{j,T}=\frac{\displaystyle\sum_{i=1}^{T}\rho_{j}X_{i-1,j}^{2}+\varepsilon_{j}(i)X_{i-1}}{\displaystyle\sum_{i=1}^{T}X_{i-1,j}^{2}}\geq
\rho_{j},\quad \mbox{a.s.},\label{A5:asbce2}
\end{equation}
\noindent for every $j\geq 1.$

From (\ref{A5:asbce2}), to obtain the following a.s. inequality:
\begin{eqnarray}
2|\widetilde{\rho}_{j,T}^{-}-\rho_{j}|&=&\left|\widehat{\rho}_{j,T}-\rho_{j}+1-\rho_{j}-\sqrt{(\widehat{\rho}_{j,T}
-1
)^{2}-\frac{4\sigma_{j}^{2}[2-(a_{j}+b_{j})]}{\beta_{j,T}}}\right|\nonumber\\
&\leq & 2|\widehat{\rho}_{j,T}-\rho_{j}| \quad
\mbox{a.s},\quad j\geq 1, \label{A5:ineqq}
\end{eqnarray}
\noindent it is sufficient that
$$-\widehat{\rho}_{j,T}+\rho_{j}\leq 1-\rho_{j}-\sqrt{(\widehat{\rho}_{j,T}-1)^{2}-\frac{4\sigma_{j}^{2}[2-(a_{j}+b_{j})]}{\beta_{j,T}}}\leq
\widehat{\rho}_{j,T}-\rho_{j} \quad \mbox{a.s},$$ \noindent
which is equivalent to
\begin{equation}
0 \leq  -\frac{2-(a_{j}+b_{j})}{\beta_{j,T}}\leq
4(\widehat{\rho}_{j,T}-\rho_{j})(1-\rho_{j})\frac{\beta_{j,T}}{4\sigma_{j}^{2}} \quad
\mbox{a.s.}.\label{A5:eqcond}
\end{equation}

That is, keeping in mind that $$\sigma_{j}^{2}=
C_j (1-\rho_{j}^{2})= C_j(1+\rho_{j})(1-\rho_{j}),$$
condition (\ref{A5:eqcond}) can also be expressed as
$$
0 \leq  -\frac{2-(a_{j}+b_{j})}{\beta_{j,T}}\leq
4(\widehat{\rho}_{j,T}-\rho_{j})(1-\rho_{j})\frac{\beta_{j,T}}{4C_j (1+\rho_{j})(1-\rho_{j})},\quad
\mbox{a.s.}
$$
\noindent i.e.,
\begin{equation}
0 \leq  -\frac{2-(a_{j}+b_{j})}{\beta_{j,T}}\leq
(\widehat{\rho}_{j,T}-\rho_{j})\frac{\beta_{j,T}}{C_j (1+\rho_{j})}\quad
\mbox{a.s}, \nonumber 
\end{equation} 
\noindent for $j\geq 1.$ Since, for each $j\geq 1,$
$$\frac{\beta_{j,T}}{C_j (1+\rho_{j})}\geq
\frac{\beta_{j,T}}{2 C_j},$$   it is
sufficient that 
\begin{equation}
0 \leq  -\frac{2-(a_{j}+b_{j})}{\beta_{j,T}}\leq
(\widehat{\rho}_{j,T}-\rho_{j})\frac{\beta_{j,T}}{2C_j}\quad
\mbox{a.s.}\label{A5:if}
\end{equation} 
\noindent to hold to ensure that  inequality (\ref{A5:ineqq}) is satisfied. Furthermore, from \textcolor{Crimson}{Remark} \ref{A5:remark2} and \textcolor{Crimson}{Corollary} \ref{A5:corollary1}, in  \textcolor{Crimson}{Supplementary Material} \ref{A5:Supp2},  for each $j\geq 1,$ $\beta_{j,T}\rightarrow
\infty,$ and $$\beta_{j,T}=\mathcal{O}(T),\quad  T\rightarrow \infty,\quad \mbox{a.s.},\quad j\geq 1.$$ \noindent Also, we have, from such remark and theorem, that $$(\widehat{\rho}_{j,T}-\rho_{j})=\mathcal{O}(1),\quad T\rightarrow \infty,\quad  \mbox{a.s.},\quad j\geq 1.$$ \noindent  Thus, for each $j\geq 1,$ the upper bound,  in (\ref{A5:if}), diverges as $T\rightarrow \infty,$ which  means, that, for $T$ sufficiently large,   inequality  (\ref{A5:ineqq}) holds, if $a_{j}+b_{j}\geq 2,$ for each $j\geq 1.$ Now, from (\ref{A5:ineqq}), under \textcolor{Aquamarine}{\textbf{Assumption A3}}, for each $j\geq 1,$

\begin{eqnarray}
T|\widehat{\rho}_{j,T}-\rho_{j}|^{2}&\leq&\widetilde{M}^{2}(j)~a.s., \quad T|\widetilde{\rho}_{j,T}^{-}-\rho_{j}|^{2} \leq T|\widehat{\rho}_{j,T}-\rho_{j}|^{2}\leq
\widetilde{M}^{2}(j)~a.s.
\label{A5:eqineqbcestv2}
\end{eqnarray}

Furthermore, for each $j\geq 1,$ $\beta_{j,T}\rightarrow
\infty,$ and $\beta_{j,T}=\mathcal{O}(T),$ as $T\rightarrow \infty,$ almost surely. Hence,
$$-\frac{4\sigma_{j}^{2}[2-(a_{j}+b_{j})]}{\beta_{j,T}}\longrightarrow
0,\quad T\longrightarrow \infty,\quad \mbox{a.s.}, \quad \forall j\geq 1.$$

From equation (\ref{A5:eqestbayes}), we then have that, for each  $j\geq 1,$
\begin{eqnarray}\lim_{T\rightarrow \infty}\left|\widetilde{\rho}_{j,T}^{-}-\widehat{\rho}_{j,T}\right|&=&
\lim_{T\rightarrow \infty}\left|\frac{1}{2}\left[(\widehat{\rho}_{j,T}+1)-
\left((\widehat{\rho}_{j,T} -1)^{2} -\frac{4}{\beta_{j,T}} \sigma_{j}^{2}[2-(a_{j}+b_{j})]\right)^{1/2}\right]- \widehat{\rho}_{j,T}\right| \nonumber \\
&=& \lim_{T\rightarrow \infty}\left|\widehat{\rho}_{j,T}- \widehat{\rho}_{j,T}\right|=
0,\nonumber\\
\label{A5:eqfconv}
\end{eqnarray}
\noindent almost surely. Thus, the almost surely convergence, when $T\rightarrow
\infty,$ of $\widetilde{\rho}_{j,T}^{-}$ and $\widehat{\rho}_{j,T}$
to the same limit is obtained, for every $j\geq 1.$

 From equation
  (\ref{A5:eqineqbcestv2}),
\begin{eqnarray}
T[\widetilde{\rho}^{-}_{j,T}-\widehat{\rho}_{j,T}]^{2}\leq 2T\left[\left(\widetilde{\rho}^{-}_{j,T}-\rho_{j}\right)^{2}+\left(\widehat{\rho}_{j,T}-\rho_{j}\right)^{2}\right]\leq 4\widetilde{M}^{2}(j),\quad
\mbox{a.s.}
\label{A5:eqdominatervbb}
\end{eqnarray}

  Since ${\rm E} \left\lbrace \widetilde{M}^{2}(j) \right\rbrace<\infty,$
 applying the  Dominated Convergence Theorem, from  equation (\ref{A5:eqdominatervbb}), considering  (\ref{A5:Bartlettf})  we obtain,    for each $j\geq 1,$
 \begin{equation}
 \lim_{T\rightarrow \infty}T {\rm E} \left\lbrace \widetilde{\rho}^{-}_{j,T}-\rho_{j} \right\rbrace^{2}=\lim_{T\rightarrow \infty}T {\rm E} \left\lbrace \widehat{\rho}_{j,T}-\rho_{j} \right\rbrace^{2}=1-\rho_{j}^{2}.\label{A5:eexmvfvf}\end{equation}

 Under  \textcolor{Aquamarine}{\textbf{Assumptions A3}}, from     (\ref{A5:eqineqbcestv2}), for each $j\geq 1,$ and  for every  $T\geq 1,$
\begin{eqnarray}
{\rm E} \left\lbrace \widehat{\rho}_{j,T}-\rho_{j} \right\rbrace^{2}\leq {\rm E} \left\lbrace \widetilde{M}^{2}(j) \right\rbrace, \quad T {\rm E} \left\lbrace \widetilde{\rho}_{j,T}^{-}-\rho_{j}\right\rbrace^{2}\leq
{\rm E} \left\lbrace \widetilde{M}^{2}(j) \right\rbrace \nonumber \end{eqnarray} 
\noindent
 with $$ \displaystyle \sum_{j=1}^{\infty } {\rm E} \left\lbrace M^{2}(j) \right\rbrace<\infty.$$
 
 Applying again the Dominated Convergence Theorem (with integration performed with respect to a counting measure), we obtain from
 (\ref{A5:eexmvfvf}), keeping in mind relationship (\ref{A5:eqreeigcrho}),
 \begin{eqnarray}
 \lim_{T\rightarrow \infty}\sum_{j=1}^{\infty}T {\rm E} \left\lbrace \widetilde{\rho}_{j,T}^{-}-\rho_{j} \right\rbrace^{2}&=&\sum_{j=1}^{\infty} \lim_{T\rightarrow \infty}T {\rm E} \left\lbrace \widetilde{\rho}_{j,T}^{-}-\rho_{j} \right\rbrace^{2} = \sum_{j=1}^{\infty} \lim_{T\rightarrow \infty}T {\rm E} \left\lbrace \widehat{\rho}_{j,T}-\rho_{j} \right\rbrace^{2} \nonumber \\
 &=&\sum_{j=1}^{\infty}
1-\rho_{j}^{2} = \sum_{j=1}^{\infty}\frac{\sigma_{j}^{2}}{C_j}=\lim_{T\rightarrow \infty}\sum_{j=1}^{\infty}
 T {\rm E} \left\lbrace \widehat{\rho}_{j,T}-\rho_{j} \right\rbrace^{2}<\infty ,\nonumber 
 \end{eqnarray}

\noindent in view of equation (\ref{A5:orderconvzeroinnv2}) in \textcolor{Aquamarine}{\textbf{Assumption A2B}}. That is, equation
(\ref{A5:eqfeqtheorem1}) holds.

\hfill \hfill \textcolor{Aquamarine}{$\blacksquare$}
\end{proof}

\bigskip

\begin{theorem}
\label{A5:mr2}  Under the conditions of \textcolor{Crimson}{Theorem} \ref{A5:mr},
\begin{eqnarray}
\lim_{T\rightarrow \infty}
T {\rm E} \left\lbrace \|\widetilde{\rho}_{T}^{-}(X_{T})-\rho
(X_{T})\|^{2}_{H} \right\rbrace = \lim_{T\rightarrow \infty}
T {\rm E} \left\lbrace \|\widehat{\rho}_{T}(X_{T})-\rho
(X_{T})\|^{2}_{H}\right\rbrace  = \sum_{k=1}^{\infty }C_k (1-\rho_{k}^{2}). \nonumber \\
\label{A5:eqasbayclarho2}
\end{eqnarray}

Here,
\begin{eqnarray}
\widetilde{\rho}_{T}^{-}(X_{T})&=&\sum_{j=1}^{\infty}\widetilde{\rho}_{j,T}^{-}\left\langle X_{T},\phi_{j}\right\rangle_{H}\phi_{j},\nonumber\\
\widetilde{\rho}_{j,T}^{-}&=&\frac{1}{2\beta_{j,T} }\left[
(\alpha_{j,T} +\beta_{j,T} )-\sqrt{(\alpha_{j,T} -\beta_{j,T}
)^{2}-4\beta_{j,T} \sigma_{j}^{2}[2-(a_{j}+b_{j})]}\right],\
j\geq
 1
\nonumber\\
\widehat{\rho}_{T}(X_{T})&=&\displaystyle \sum_{j=1}^{\infty}\widehat{\rho}_{j,T}\left\langle
X_{T},\phi_{j}\right\rangle_{H}\phi_{j},\quad
\widehat{\rho}_{j,T}\frac{\displaystyle \sum_{i=1}^{T}X_{i-1,j}X_{i,j}}{\displaystyle \sum_{i=1}^{T}X_{i-1,j}^{2}},\quad
j\geq
 1\nonumber\\
 \rho
(X_{T})&=&\sum_{j=1}^{\infty}\rho_{j}\left\langle
X_{T},\phi_{j}\right\rangle_{H}\phi_{j},\quad
\rho_{j}=\rho(\phi_{j})(\phi_{j}),\quad j\geq 1. \nonumber
\end{eqnarray}
\end{theorem}

\begin{proof}

From equation (\ref{A5:eqfconv}), for every $j,k\geq 1,$\begin{equation}
\left[\left(\widetilde{\rho}^{-}_{j,T}-\widehat{\rho}_{j,T} \right)
\left(\widetilde{\rho}^{-}_{k,T}-\widehat{\rho}_{k,T}\right) \right]^{2}\rightarrow 0,\quad \mbox{a.s.},\quad T\rightarrow \infty.
\label{A5:eqasth1}
\end{equation}

In addition, from equation (\ref{A5:eqdominatervbb}),  for every $j,k\geq 1,$
\begin{equation}
\left[ \left(\widetilde{\rho}^{-}_{j,T}-\widehat{\rho}_{j,T} \right)
\left(\widetilde{\rho}^{-}_{k,T}-\widehat{\rho}_{k,T}\right)\right]^{2}\leq 16 \frac{\widetilde{M}^{2}(k)\widetilde{M}^{2}(j)}{T^{2}}\leq
16\widetilde{M}^{2}(k)\widetilde{M}^{2}(j),
\label{A5:eq2th2}
\end{equation}
\noindent with $${\rm E} \left\lbrace \widetilde{M}^{2}(k)\widetilde{M}^{2}(j) \right\rbrace = {\rm E} \left\lbrace \widetilde{M}^{2}(k) \right\rbrace {\rm E} \left\lbrace \widetilde{M}^{2}(j) \right\rbrace<\infty,$$
under \textcolor{Aquamarine}{\textbf{Assumption A3}}. Applying the Dominated Convergence Theorem from (\ref{A5:eq2th2}), the almost surely convergence in (\ref{A5:eqasth1}) implies
the convergence in mean to zero, when $T\rightarrow \infty.$  Furthermore, under \textcolor{Aquamarine}{\textbf{Assumption A3}}, for $T\geq 2,$
\begin{eqnarray}
\sum_{j=1}^{\infty}\sum_{k=1}^{\infty}T^{2}{\rm E} \left\lbrace \left(\widetilde{\rho}^{-}_{j,T}-\widehat{\rho}_{j,T} \right) \left(\widetilde{\rho}^{-}_{k,T}-\widehat{\rho}_{k,T} \right) \right]^{2} &\leq & 16\left[\sum_{\substack{j,k=1 \\ j\neq k}}^{\infty} {\rm E} \left\lbrace \widetilde{M}^{2}(j) \right\rbrace {\rm E} \left\lbrace \widetilde{M}^{2}(k) \right\rbrace \right]
 \nonumber \\
 &+& \left[\sum_{k=1}^{\infty } {\rm E} \left\lbrace \widetilde{M}^{4}(k) \right\rbrace\right]<\infty.\label{A5:eq2th3}
\end{eqnarray}

From (\ref{A5:eq2th3}), for every $T\geq 2,$
\begin{eqnarray}
T^{2} {\rm E} \left\lbrace \|\widetilde{\rho}^{-}_{T}-\widehat{\rho}_{T}\|_{\mathcal{S}(H)}^{4} \right\rbrace
&=& \sum_{j=1}^{\infty}\sum_{k=1}^{\infty}T^{2} {\rm E} \left\lbrace (\widetilde{\rho}^{-}_{j,T}-\widehat{\rho}_{j,T})
(\widetilde{\rho}^{-}_{k,T}-\widehat{\rho}_{k,T}) \right\rbrace^{2}\nonumber\\
 &\leq & 16\left[\sum_{\substack{j,k = 1 \\ j\neq k}}^{\infty}{\rm E} \left\lbrace \widetilde{M}^{2}(j) \right\rbrace {\rm E} \left\lbrace \widetilde{M}^{2}(k) \right\rbrace \right] \nonumber \\
 &+& \left[\sum_{k=1}^{\infty }{\rm E} \left\lbrace \widetilde{M}^{4}(k) \right\rbrace \right]<\infty.
\label{A5:e2th4}
\end{eqnarray}

Equation (\ref{A5:e2th4}) means that the rate of convergence to zero, as $T\rightarrow \infty,$  of the functional sequence \linebreak $\left\lbrace \widetilde{\rho}^{-}_{T}-\widehat{\rho}_{T}, \ T\geq 2 \right\rbrace$ in the space  $\mathcal{L}_{\mathcal{S}(H)}^{4}(\Omega,\mathcal{A},P)$ is of order $T^{-2}.$

From definition of the norm in the space bounded linear operators, applying  the Cauchy--Schwarz's inequality, we obtain
 \begin{eqnarray}
{\rm E} \left\lbrace \|\widetilde{\rho}_{T}^{-}(X_{T})-\widehat{\rho}_{T}
(X_{T})\|_{H}^{2} \right\rbrace &\leq & {\rm E} \left\lbrace \|\widetilde{\rho}_{T}^{-}-\widehat{\rho}_{T}\|_{\mathcal{L}(H)}^{2}\|X_{T}\|_{H}^{2}
\right\rbrace \nonumber\\
&\leq & \sqrt{ {\rm E} \left\lbrace \|\widetilde{\rho}_{T}^{-}-\widehat{\rho}_{T}\|_{\mathcal{L}(H)}^{4}\right\rbrace}\sqrt{ {\rm E} \left\lbrace \|X_{T}\|_{H}^{4}\right\rbrace}
\nonumber\\
&\leq &\sqrt{ {\rm E} \left\lbrace \|\widetilde{\rho}_{T}^{-}-\widehat{\rho}_{T}\|_{\mathcal{S}(H)}^{4}\right\rbrace}\sqrt{{\rm E} \left\lbrace \|X_{T}\|_{H}^{4}\right\rbrace}.
\label{A5:tr2eq5}
\end{eqnarray}

From the orthogonal expansion (\ref{A5:kesxp}) of $X_{T}$, in terms of the independent real--valued standard Gaussian random variables  $\left\lbrace \eta_{k}(T), \ k\geq 1 \right\rbrace,$ we have
 \begin{eqnarray}
 {\rm E} \left\lbrace \|X_{T}\|_{H}^{4}\right\rbrace &=&\sum_{j=1}^{\infty}\sum_{k=1}^{\infty}C_j C_k {\rm E} \left\lbrace \eta_{j}(T)\eta_{k}(T) \right\rbrace ^{2}=\sum_{j=1}^{\infty}\sum_{k=1}^{\infty}C_j C_k 3\delta_{j,k} \nonumber \\
 &=& 3\sum_{k=1}^{\infty}C_{j}^{2} <\infty.
\label{A5:tr2eq6}
\end{eqnarray}

From equations (\ref{A5:e2th4})--(\ref{A5:tr2eq6}), $${\rm E} \left\lbrace \|\widetilde{\rho}_{T}^{-}(X_{T})-\widehat{\rho}_{T}
(X_{T})\|_{H}^{2} \right\rbrace =\mathcal{O}\left(\frac{1}{T}\right),\quad T\rightarrow \infty.$$

Thus,
$\widetilde{\rho}_{T}^{-}(X_{T})$ and $\widehat{\rho}_{T}
(X_{T})$ have the same limit in the space $\mathcal{L}_{H}^{2}(\Omega,\mathcal{A},\mathcal{P}).$

\bigskip

We now prove the approximation by ${\rm Tr} \left(C \left(I-\rho^{2} \right)\right)$ of the  limit, in equation   (\ref{A5:eqasbayclarho2}).
Consider
\begin{equation}
{\rm E} \left\lbrace \|\widehat{\rho}_{T}(X_{T})-\rho
(X_{T})\|^{2}_{H} \right\rbrace - {\rm Tr} \left(C(I-\rho^{2})\right) = 
\sum_{k=1}^{\infty} {\rm E} \left\lbrace \left(\widehat{\rho}_{k,T}-\rho_{k}\right)^{2}\eta_{k}^{2}(T) \right\rbrace C_k - C_k(1-\rho_{k}^{2}),
\label{A5:trcelim}
\end{equation}
\noindent where
$${\rm Tr} \left(C(I-\rho^{2})\right)=\sum_{k=1}^{\infty} C_k (1-\rho_{k}^{2}).$$

From \textcolor{Crimson}{Lemmas} \ref{A5:prdistab}-- \ref{A5:lem1}
 (see the last identity in equation (\ref{A5:eqmomentnrvs})), for each $k\geq 1,$ and for $T$ sufficiently large,
\begin{eqnarray}
{\rm E} \left\lbrace \left(\widehat{\rho}_{k,T}-\rho_{k} \right)^{2}\eta_{k}^{2}(T)\right\rbrace &\simeq & {\rm Var} \left\lbrace \widehat{\rho}_{k,T}-\rho_{k} \right\rbrace {\rm Var} \left\lbrace \eta_{k} \right\rbrace
 \times \left( 1+2\left[{\rm Corr} \left( \widehat{\rho}_{k,T}-\rho_{k},\eta_{k}(T) \right) \right]^{2}\right). \nonumber \\ \label{A5:id1}
\end{eqnarray}

Under \textcolor{Aquamarine}{\textbf{Assumption A3}}, from equations (\ref{A5:asi22})--(\ref{A5:eqest}), for every $k\geq 1,$

\begin{equation}
T {\rm Var} \left\lbrace \widehat{\rho}_{k,T}-\rho_{k} \right\rbrace \leq \left(1-\rho_{k}^{2}\right) {\rm E} \left\lbrace \widetilde{M}^{2}(k) \right\rbrace. \label{A5:varrhokest}
\end{equation}

 From equations (\ref{A5:trcelim})--(\ref{A5:varrhokest}),
\begin{eqnarray}
T {\rm E} \left\lbrace \|\widehat{\rho}_{T}(X_{T})-\rho
(X_{T})\|^{2}_{H} \right\rbrace - {\rm Tr} \left(C(I-\rho^{2})\right) & \leq &
\sum_{k=1}^{\infty}C_k (1-\rho_{k}^{2}) {\rm E} \left\lbrace \widetilde{M}^{2}(k) \right\rbrace
\nonumber\\
& \times &
\left[1+2\left[{\rm Corr} \left( \widehat{\rho}_{k,T}-\rho_{k},\eta_{k}(T)\right)\right]^{2}\right] \nonumber \\
&-& C_k(1-\rho_{k}^{2}) \leq 
\sum_{k=1}^{\infty}3C_k {\rm E} \left\lbrace \widetilde{M}^{2}(k) \right\rbrace \nonumber \\
&-&\sum_{k=1}^{\infty}C_k (1-\rho_{k}^{2})<\infty,
\label{A5:finit}
\end{eqnarray}
\noindent since $$
\sum_{k=1}^{\infty}C_k(1-\rho_{k}^{2})\leq
\sum_{k=1}^{\infty}C_k< \infty,$$ by the trace property of
$C.$ Here, we have  applied   the Cauchy--Schwarz's inequality to obtain,
for a certain constant  $L>0,$
\begin{eqnarray}\sum_{k=1}^{\infty}3C_k {\rm E} \left\lbrace \widetilde{M}^{2}(k) \right\rbrace &\leq & 3\sqrt{\sum_{k=1}^{\infty
}C_{k}^{2} \sum_{k=1}^{\infty}\left[{\rm E} \left\lbrace \widetilde{M}^{2}(k) \right\rbrace \right]^{2}}\nonumber\\
&\leq & 3L\sqrt{\sum_{k=1}^{\infty
}C_k \sum_{k=1}^{\infty} {\rm E} \left\lbrace \widetilde{M}^{2}(k) \right\rbrace}<\infty, \nonumber 
\end{eqnarray} \noindent from the trace property of $C,$  and since  $$\sum_{k=1}^{\infty}{\rm E} \left\lbrace \widetilde{M}^{2}(k)\right\rbrace <\infty,$$ under \textcolor{Aquamarine}{\textbf{Assumption A3}}.

 From equations (\ref{A5:Bartlettf}) and
(\ref{A5:finit}), one can get, applying the Dominated Convergence Theorem,
\begin{eqnarray}
\lim_{T\rightarrow \infty}T {\rm E} \left\lbrace \|\widehat{\rho}_{T}(X_{T})-\rho
(X_{T})\|^{2}_{H} \right\rbrace &=& \displaystyle \sum_{k=1}^{\infty}
C_k \displaystyle \lim_{T\rightarrow \infty}T {\rm E} \left\lbrace \widehat{\rho}_{k,T}-\rho_{k}\right\rbrace^{2}\nonumber\\
& \times &
\displaystyle \lim_{T\rightarrow \infty}\left[1+\left[{\rm Corr} \left( \widehat{\rho}_{k,T}-\rho_{k},\eta_{k}(T) \right) \right]^{2}\right]\nonumber\\
&=& \sum_{k=1}^{\infty} C_k \lim_{T\rightarrow
\infty}T {\rm E} \left\lbrace \widehat{\rho}_{k,T}-\rho_{k}\right\rbrace^{2} \nonumber \\
& = & \displaystyle \sum_{k=1}^{\infty} C_k (1-\rho_{k}^{2}), \nonumber 
\end{eqnarray}
\noindent where we have considered that
\begin{eqnarray}
\displaystyle \lim_{T\rightarrow
\infty} \left|{\rm Cov} \left( \widehat{\rho}_{k,T}-\rho_{k},\eta_{k}(T) \right) \right|^{2} &\leq &
\displaystyle \lim_{T\rightarrow
\infty} {\rm E} \left\lbrace \widehat{\rho}_{k,T}-\rho_{k} \right\rbrace^{2} {\rm E} \left\lbrace \eta_{k}(T) \right\rbrace^{2} = \displaystyle \lim_{T\rightarrow \infty}\frac{1-\rho^{2}_{k}}{T}=0.\nonumber
\end{eqnarray}

\hfill \hfill \textcolor{Aquamarine}{$\blacksquare$}
\end{proof}

%
%

\textcolor{Crimson}{\section{Numerical examples}}
\label{A5:sim_study_}
This section  illustrates the theoretical results derived on asymptotic efficiency and equivalence of the proposed
classical and Bayesian diagonal  componentwise  estimators of the autocorrelation operator, as well as of the associated ARH(1) plug--in predictors.
Under the conditions assumed in \textcolor{Crimson}{Theorem} \ref{A5:mr},
three examples of
  standard zero--mean Gaussian ARH(1) processes  are generated, respectively  corresponding to consider  different rates of convergence to zero of the  eigenvalues  of the autocovariance operator. The truncation order $k_{T}$ in \textcolor{Crimson}{Examples 1--2} (see \textcolor{Crimson}{Sections}   \ref{A5:Ex1}--\ref{A5:Ex2}) is fixed; i.e., it does not depend on the sample size $T$ (see equations (\ref{A5:61})--(\ref{A5:62}) below). While in \textcolor{Crimson}{Example 3} (see \textcolor{Crimson}{Section} \ref{A5:Ex3}), $k_{T}$ is selected such that 
  \begin{equation}\lim_{T\rightarrow \infty}C_{k_{T}}\sqrt{T}=\infty.\label{trorderrule}
  \end{equation}
  
  Specifically, in the first two examples, the choice of $k_{T}$  is driven  looking for a compromise between the sample size and the number of parameters to be estimated. With this aim the value  $k_T = 5$  is fixed,  independently  of $T.$    This is the number of parameters that can be estimated
     in an efficient way, from most of the values of the sample size $T$ studied. In \textcolor{Crimson}{Example 3}, the truncation parameter $k_{T}$ is defined as a fractional power of the sample size. Note that \textcolor{Crimson}{Example 3} corresponds to  the fastest decay velocity of the eigenvalues of the autocovariance operator. Hence, the lowest truncation order for a given sample size must be selected according to the truncation rule (\ref{trorderrule}).

 The
generation of $N=1000$ realizations of the functional values $\left\lbrace X_{t}, \ t=0,1,\dots, T \right\rbrace$, for  $$T= \left[250,  500, 750, 1000, 1250, 1500, 1750, 2000\right],$$ denoting as before the sample size, is performed, for each one of the ARH(1) processes, defined  in the three examples  below.
 Based on those generations, and on the sample sizes studied, the truncated empirical functional mean-square
errors of the classical and    Bayes  diagonal componentwise parameter
estimators of the autocorrelation operator $\rho$  are computed as follows:

\begin{eqnarray}
{\rm EFMSE}_{\overline{\rho}_T} &=& \frac{1}{N} \displaystyle \sum_{\omega=1}^{N} \displaystyle \sum_{j=1}^{k_n} \left(\overline{\rho}_{j,T}^{\omega} - \rho_j \right)^2, \label{A5:61} \\
{\rm EFMSE}_{\overline{\rho}_T \left(X_{T} \right)} &=& \frac{1}{N} \displaystyle \sum_{\omega=1}^{N} \displaystyle \sum_{j=1}^{k_n} \left(\overline{\rho}_{j,T}^{\omega} - \rho_j \right)^2 X_{T,j}^{2}, \label{A5:62}
\end{eqnarray}
\noindent where $\overline{\rho}_{j,T}^{\omega}$ can be the classical
$\widehat{\rho}_{j,T}$ or the  Bayes $\widetilde{\rho}_{j,T}$ diagonal  componentwise estimator of the autocorrelation operator, and $^{\omega}$ denotes the sample point $\omega \in \Omega $ associated with each one of the $N=1000$ realizations generated of each functional value of the ARH(1) process $X.$

On the other hand, as assumed in the previous section, $$\rho_k \sim \mathcal{B} \left(a_k, b_k \right), \quad a_k + b_k \geq 2, \quad a_k > 0, \quad b_k > 1,$$ for each $k \geq 1$. Thus, parameters $\left(a_k, b_k \right)$ are defined as follows:
\begin{equation}
b_k = 1 + 1/100, \quad a_k = 2^k, \quad k \geq 1, \label{A5:53}
\end{equation}
\noindent where
\begin{equation}
{\rm E} \left\lbrace \rho_k \right\rbrace = \frac{a_k}{a_k + b_k} \to 1, \quad {\rm Var} \left\lbrace \rho_k \right\rbrace = \frac{a_k  b_k}{\left(a_k + b_k + 1 \right) \left(a_k + b_k \right)^2} = \mathcal{O} \left(\frac{1}{ 2^{2k}} \right), \quad k \rightarrow \infty, \label{A5:54}
\end{equation}
\noindent with $\left\lbrace \rho_{k}^{2}, \ k \geq 1 \right\rbrace$ being  a random sequence such that its elements tend to be concentrated around point one, when $k\rightarrow \infty.$  From (\ref{A5:54}), since
\begin{equation}
\sigma_{k}^{2} = C_k \left( 1 - \rho_{k}^{2} \right), \quad k \geq 1, \label{A5:55}
\end{equation}

\textcolor{Aquamarine}{\textbf{Assumption A2B}} is satisfied.
In addition, condition (\ref{A5:eq2th1}) is verified in the generations performed in  the Gaussian framework.

\textcolor{Crimson}{\subsection{Example 1}
\label{A5:Ex1}}

Let us assume that the eigenvalues of the autocovariance operator of the ARH(1) process $X$ are
given by
\begin{eqnarray}
C_k &=& \frac{1}{k^{3/2}},\quad k\geq 1.\nonumber
\end{eqnarray}

Thus,  $C$ is a strictly positive and trace operator, where $$\left\lbrace \rho_{k}^{2}, \ k \geq 1 \right\rbrace, \quad \left\lbrace \sigma_{k}^{2}, \ k \geq 1 \right\rbrace,$$ are generated from (\ref{A5:53})--(\ref{A5:55}).

\begin{figure}[H]
\centering
\includegraphics[width=0.8\textwidth]{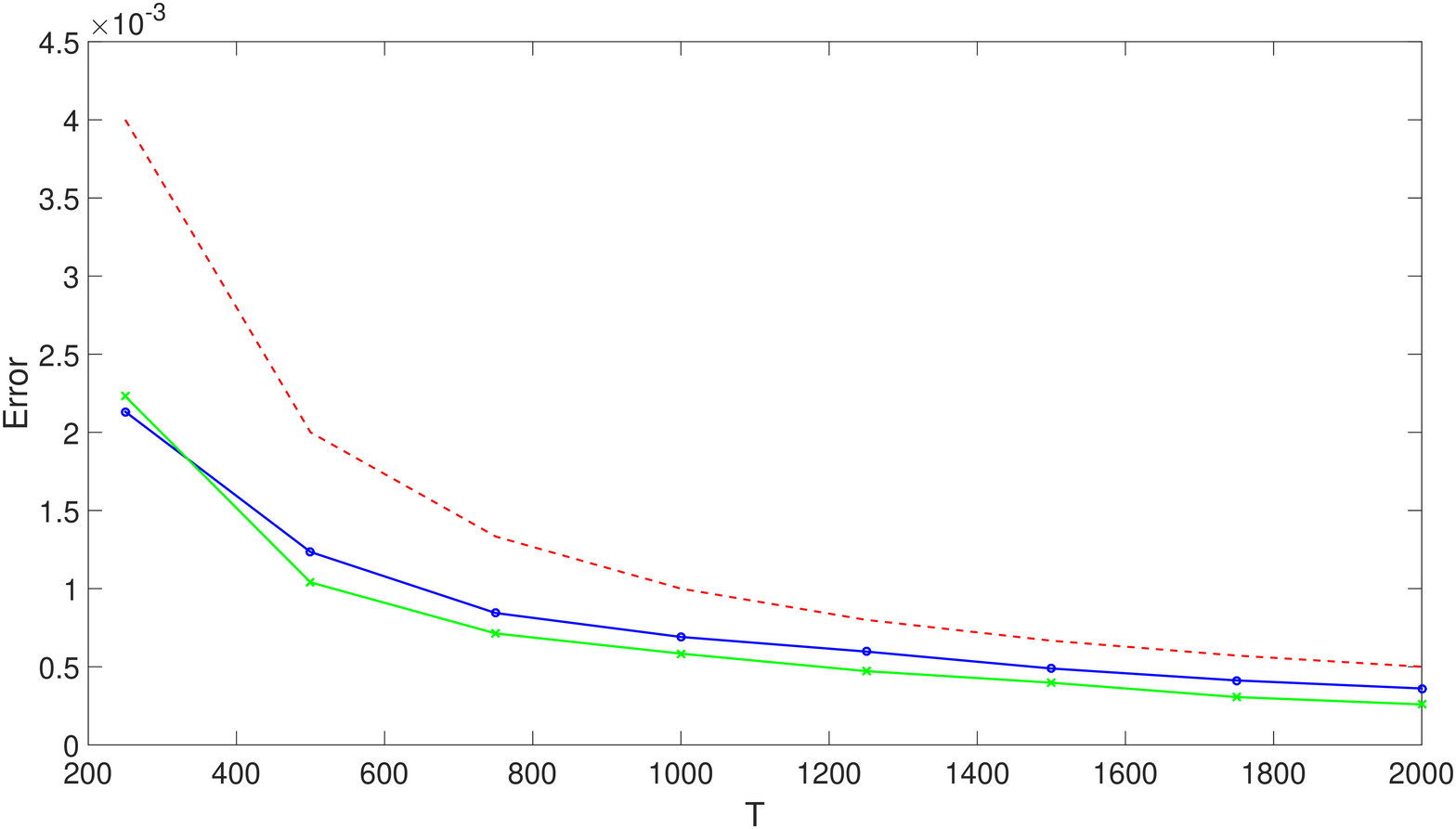}

\vspace{-0.25cm}
\caption[\hspace{0.7cm} Empirical quadratic functional errors for Example 1.]{\small{\textcolor{Crimson}{Example 1.}  Empirical functional mean-square estimation errors of
classical (blue circle line), and  Bayes (green cross line) componentwise ARH(1)
parameter estimators,  with  $k_T = 5,$ for $N=1000$ replications of the ARH(1) values,
 against the curve $1/T$ (red dot line), for  $T= \left[250,  500, 750, 1000, 1250, 1500, 1750, 2000\right]$.}}\label{A5:fig:1}
\end{figure}

\begin{figure}[H]
\centering
\includegraphics[width=0.8\textwidth]{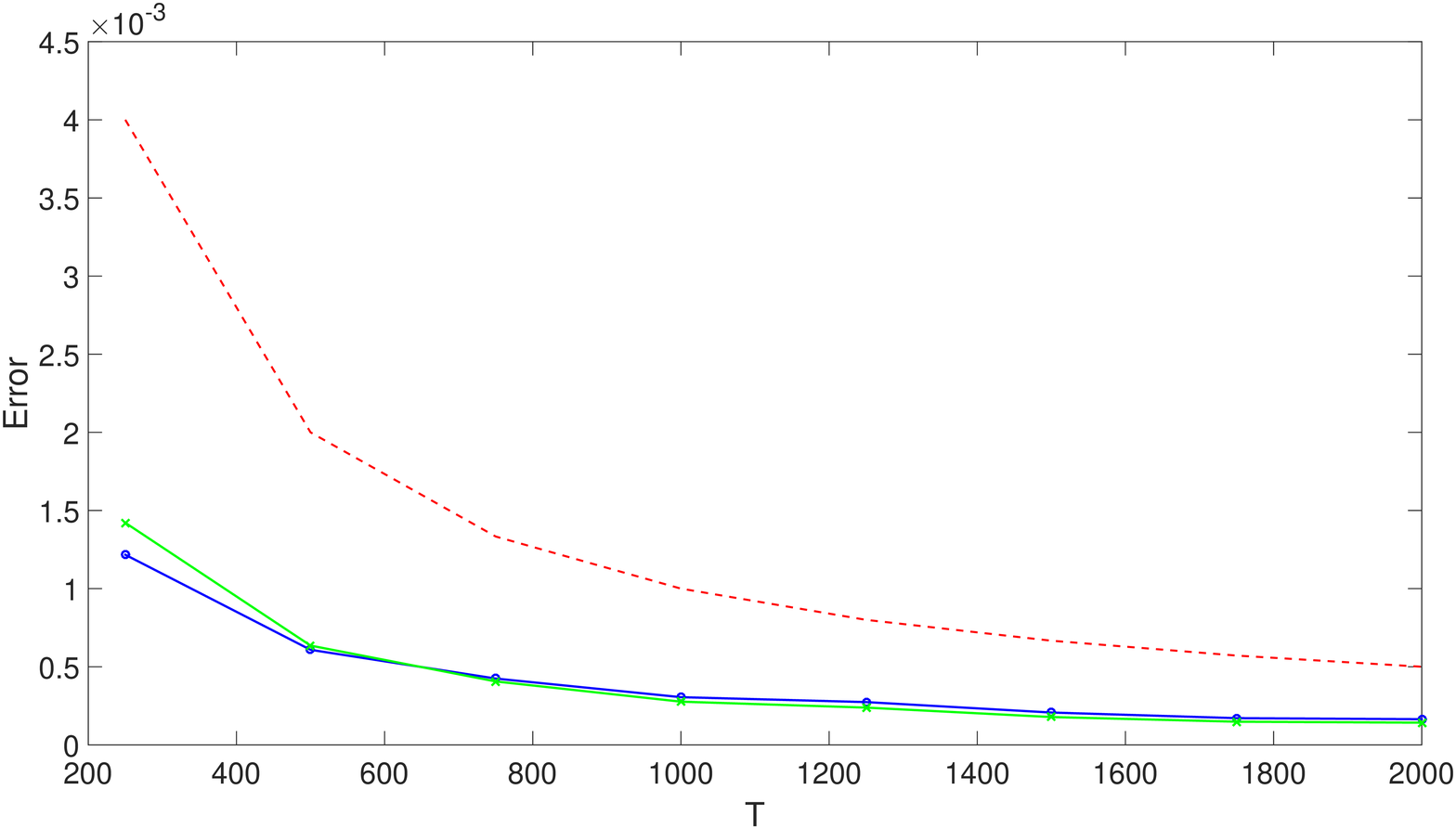}

\vspace{-0.25cm}
\caption[\hspace{0.7cm} Empirical quadratic functional errors for Example 1.]{\small{\textcolor{Crimson}{Example 1.}  Empirical functional mean-square prediction errors of
classical (blue circle line), and  Bayes (green cross line) componentwise ARH(1)
plug-in predictors, with $k_T = 5,$ for $N=1000$ replications of the ARH(1) values,
 against the curve $1/T$ (red dot line), for  $T= \left[ 250,  500, 750, 1000, 1250, 1500, 1750, 2000 \right]$.}}\label{A5:fig:2}
\end{figure}

Tables \ref{A5:tab1}--\ref{A5:tab2}  display the  values of the empirical
functional mean--square errors, given in (\ref{A5:61})--(\ref{A5:62}),   associated with  $\widehat{\rho }_{T}$ and
 $\widetilde{\rho}^{-}_{T},$ and with   the  corresponding  ARH(1)
plug--in  predictors, with, as before,  \begin{equation}T=\left[250, 500,
750, 1000, 1250, 1500, 1750, 2000 \right],\label{A5:ss}
\end{equation} \noindent considering  $k_T = 5.$   The respective graphical
representations are  displayed in Figures \ref{A5:fig:1}--\ref{A5:fig:2}, where,  for comparative purposes,
the values of the
curve $1/T$ are also drawn for the finite
sample sizes (\ref{A5:ss}).

\begin{table}[H]
\caption[\hspace{0.7cm} Empirical quadratic functional errors for \textcolor{Crimson}{Example 1} (with a fixed truncation parameter).]{\small{\textcolor{Crimson}{Example 1.} Empirical functional mean-square errors $EFMSE_{\overline{\rho}_{T}}$.}}
\label{A5:tab1}
\centering
\begin{small}
\begin{tabular}{|c|c|c|}
\hline Sample size  & Classical   estimator $\widehat{\rho}_{T}$ & Bayes estimator $\widetilde{\rho}^{-}_{T}$\\
\hline 
\hline
$250$ & $2.13\left(10 \right)^{-3}$ &  $2.23\left(10 \right)^{-3}$\\
\hline
 $500$ & $1.24\left(10 \right)^{-3}$& $1.04\left(10 \right)^{-3}$\\
 \hline
 $750$ & $8.44 \left(10 \right)^{-4}$ & $7.13\left(10 \right)^{-4}$\\
 \hline
 $1000$ & $6.91 \left(10 \right)^{-4}$  & $5.84\left(10 \right)^{-4}$\\
 \hline
 $1250$ &$ 5.97 \left(10 \right)^{-4}$ & $4.72\left(10 \right)^{-4}$\\
 \hline
 $1500$ & $4.89 \left(10 \right)^{-4}$ & $3.98 \left(10 \right)^{-4}$\\
 \hline
 $1750$ & $4.13 \left(10 \right)^{-4}$ & $3.06 \left(10 \right)^{-4}$\\
 \hline
 $2000$ & $3.61 \left(10 \right)^{-4}$ & $2.59\left(10 \right)^{-4}$\\
 \hline
\end{tabular}
\end{small}
\end{table}

\begin{table}[H]
\caption[\hspace{0.7cm} Empirical quadratic functional errors for \textcolor{Crimson}{Example 1} (with a fixed truncation parameter).]{\small{\textcolor{Crimson}{Example 1.} Empirical functional mean-square errors $EFMSE_{\overline{\rho}_{T} \left(X_T \right)}$.}}
\label{A5:tab2}
\centering
\begin{small}
\begin{tabular}{|c|c|c|}
\hline 
Sample size  & Classical  predictor $\widehat{\rho}_{T} \left(X_T \right)$ & Bayes
predictor $\widetilde{\rho}^{-}_{T}\left(X_T \right)$\\
\hline 
\hline
$250$ & $1.22 \left(10 \right)^{-3}$ & $1.42 \left(10 \right)^{-3}$\\
\hline 
 $500$ & $6.08 \left(10 \right)^{-4}$&  $6.36 \left(10 \right)^{-4}$\\
 \hline 
 $750$ & $3.24\left(10 \right)^{-4}$&  $4.06 \left(10 \right)^{-4}$\\
 \hline 
 $1000$ & $3.05 \left(10 \right)^{-4}$ & $2.77 \left(10 \right)^{-4}$\\
 \hline 
 $1250$ & $2.74 \left(10 \right)^{-4}$ &  $2.39 \left(10 \right)^{-4}$\\
 \hline 
 $1500$ & $2.07 \left(10 \right)^{-4}$ & $1.78  \left(10 \right)^{-4}$\\
 \hline 
 $1750$ & $1.71 \left(10 \right)^{-4}$ &  $1.48 \left(10 \right)^{-4}$\\
 \hline 
 $2000$ & $1.64 \left(10 \right)^{-4}$ & $1.42 \left(10 \right)^{-4}$\\
 \hline
\end{tabular}
\end{small}
\end{table}

\textcolor{Crimson}{\subsection{Example 2} \label{A5:Ex2}}

In this example, a bit slower decay velocity, than in \textcolor{Crimson}{Example 1}, of
the eigenvalues of the autocovariance operator of the ARH(1)
process is considered. Specifically,
\begin{eqnarray}
C_k &=& \frac{1}{k^{1+1/10}},\quad k \geq 1\nonumber.
\end{eqnarray}

Thus, $C$ is a strictly positive   self-adjoint trace operator, where $\left\lbrace \rho_{k}^{2}, \ k \geq 1 \right\rbrace$ and $\left\lbrace \sigma_{k}^{2}, \ k \geq 1 \right\rbrace$ are generated, as before,  from (\ref{A5:53})-(\ref{A5:55}).

\begin{figure}[H]
\centering
\includegraphics[width=0.8\textwidth]{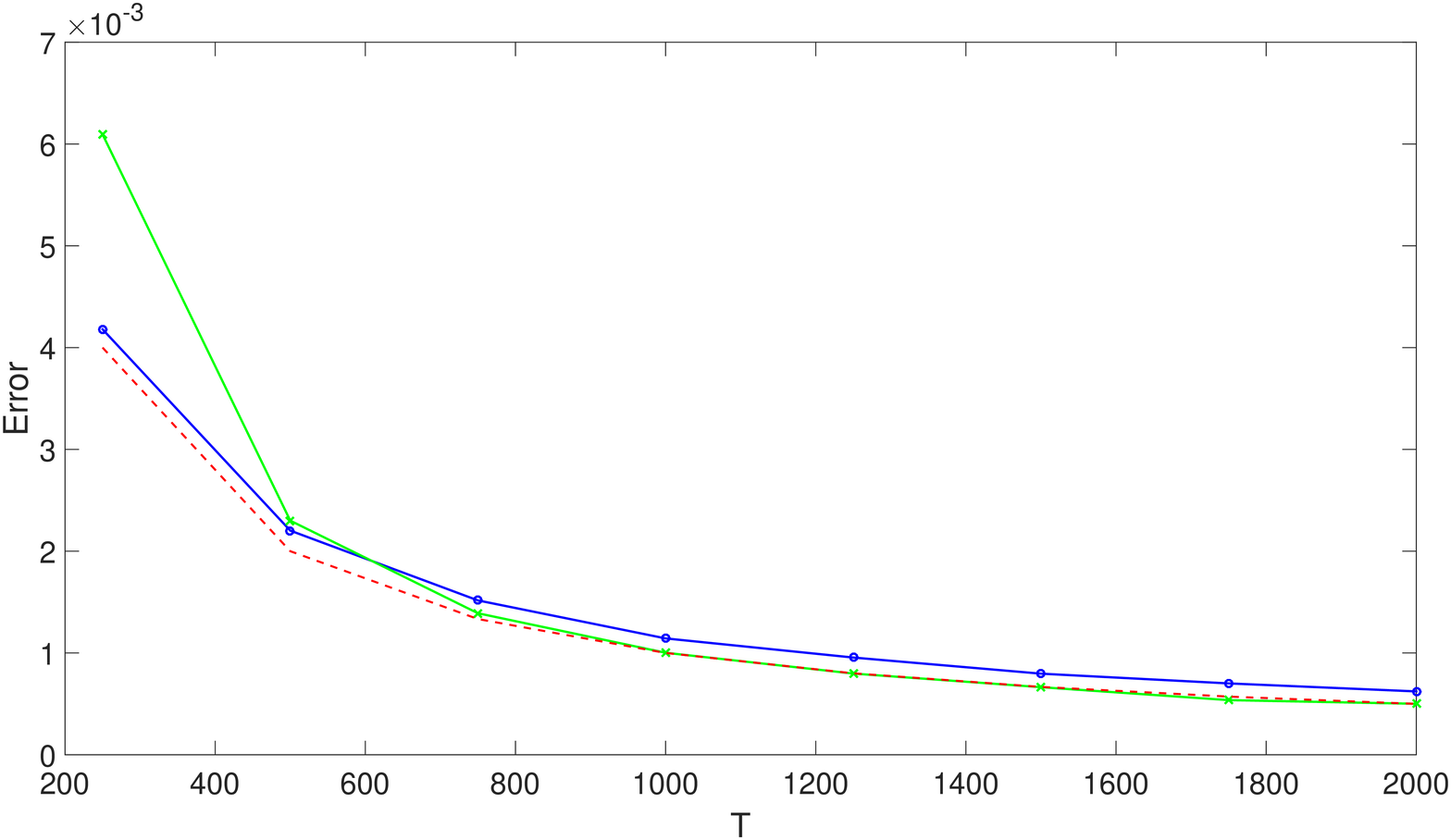}

\vspace{-0.25cm}
\caption[\hspace{0.7cm} Empirical quadratic functional errors for Example 2.]{\small{\textcolor{Crimson}{Example 2.}   Empirical functional mean--square estimation errors of
classical (blue circle line), and  Bayes (green cross line) componentwise ARH(1)
parameter estimators,  with $k_T = 5,$ for $N=1000$ replications of the ARH(1) values,
 against the curve $1/T$ (red dot line), for \linebreak $T=\left[ 250,  500, 750, 1000, 1250, 1500, 1750, 2000 \right].$}}\label{A5:fig3}
\end{figure}

Tables \ref{A5:tab4}--\ref{A5:tab5}  show  the values of
the empirical functional mean--square errors, associated with
$\widehat{\rho}_{T}$ and  $\widetilde{\rho}^{-}_{T},$ and  with
the corresponding ARH(1) plug--in  predictors, respectively. Figures \ref{A5:fig3}--\ref{A5:fig4} provide the
 graphical representations in comparison with the values
of the curve $1/T$ for   $T$ given in  (\ref{A5:ss}),  with, as before,  $k_T = 5$.

\begin{table}[H]
\caption[\hspace{0.7cm} Empirical quadratic functional errors for \textcolor{Crimson}{Example 2} (with a fixed truncation parameter a slower decay rate of eigenvalues).]{\small{\textcolor{Crimson}{Example 2.} Empirical functional mean--square errors $EFMSE_{\overline{\rho}_{T}}$.}}
\label{A5:tab4}
\centering
\begin{small}
\begin{tabular}{|c|c|c|}
\hline Sample size  & Classical  estimator $\widehat{\rho}_{T}$ & Bayes
 estimator $\widetilde{\rho}^{-}_{T}$\\
\hline 
$250$ & $4.18 \left(10 \right)^{-3}$& $6.09 \left(10 \right)^{-3}$ \\
\hline
 $500$ & $2.20 \left(10 \right)^{-3}$& $2.30 \left(10 \right)^{-3}$ \\
 \hline
 $750$ & $1.52 \left(10 \right)^{-3}$ & $1.39 \left(10 \right)^{-3}$\\
 \hline
 $1000$ & $1.14 \left(10 \right)^{-3}$  & $1.00 \left(10 \right)^{-3}$\\
 \hline
 $1250$ & $9.55 \left(10 \right)^{-4}$&  $7.97\left(10 \right)^{-4}$\\
 \hline
 $1500$ & $7.97 \left(10 \right)^{-4}$ & $6.64 \left(10 \right)^{-4}$\\
 \hline
 $1750$ & $7.01 \left(10 \right)^{-4}$ & $5.37 \left(10 \right)^{-4}$\\
 \hline
 $2000$ & $6.22 \left(10 \right)^{-4}$ & $5.00\left(10 \right)^{-4}$\\
 \hline
\end{tabular}
\end{small}
\end{table}

\begin{table}[H]
\caption[\hspace{0.7cm} Empirical quadratic functional errors for \textcolor{Crimson}{Example 2} (with a fixed truncation parameter and slower decay rate of eigenvalues).]{\small{\textcolor{Crimson}{Example 2.} Empirical functional mean--square errors $EFMSE_{\overline{\rho}_{T} \left(X_T \right)}$.}}
\label{A5:tab5}
\centering
\begin{small}
\begin{tabular}{|c|c|c|}
\hline Sample size   & Classical  predictor $\widehat{\rho}_{T} \left(X_T \right)$ & Bayes
predictor $\widetilde{\rho}^{-}_{T}\left(X_T \right)$\\
\hline 
$250$ & $3.25 \left(10 \right)^{-3}$ & $3.18 \left(10 \right)^{-4}$\\
\hline
 $500$ & $1.59 \left(10 \right)^{-3}$ & $ 1.40\left(10 \right)^{-4}$\\
 \hline
 $750$ & $9.47 \left(10 \right)^{-4}$ &  $8.19 \left(10 \right)^{-4}$\\
 \hline
 $1000$ & $7.89 \left(10 \right)^{-4}$&$ 6.88 \left(10 \right)^{-4}$\\
 \hline
 $1250$ & $7.24\left(10 \right)^{-4}$&  $6.10 \left(10 \right)^{-4}$\\
 \hline
 $1500$ & $5.53\left(10 \right)^{-4}$& $4.77 \left(10 \right)^{-4}$\\
 \hline
 $1750$ & $5.31 \left(10 \right)^{-4}$ &  $4.49\left(10 \right)^{-4}$\\
 \hline
 $2000$ & $4.61 \left(10 \right)^{-4}$& $4.00\left(10 \right)^{-4}$\\
 \hline
\end{tabular}
\end{small}
\end{table}

\begin{figure}[H]
\label{A5:fig4}
\centering
\includegraphics[width=0.8\textwidth]{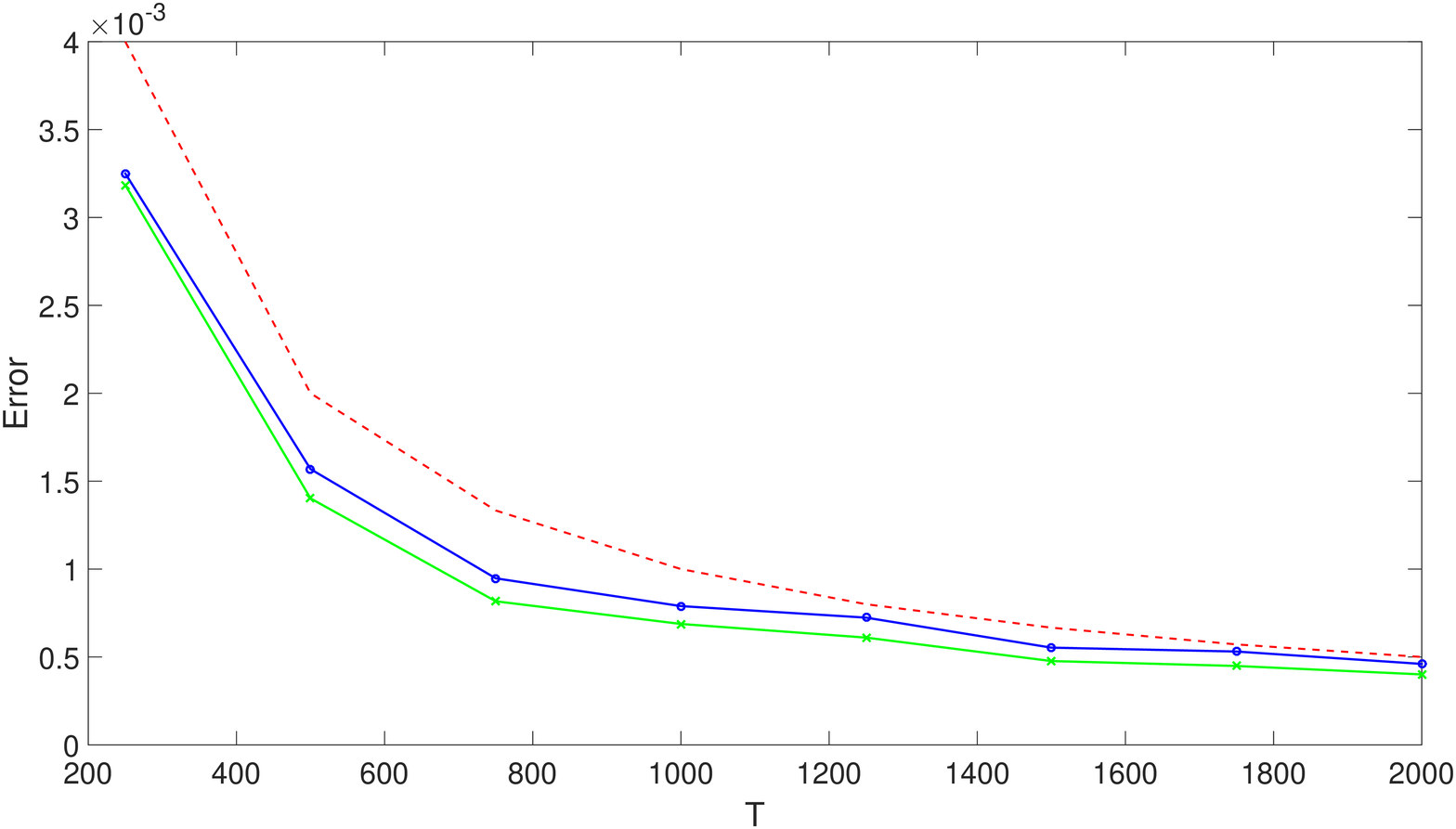}

\vspace{-0.25cm}
\caption[\hspace{0.7cm} Empirical quadratic functional errors for Example 2.]{\small{\textcolor{Crimson}{Example 2.}   Empirical functional mean--square prediction errors of
classical (blue circle line), and  Bayes (green cross line) componentwise ARH(1)
plug-in predictors,  with $k_T = 5,$ for $N=1000$ replications of the ARH(1) values,
 against the curve $1/T$ (red dot line), for  \linebreak $T= \left[ 250,  500, 750, 1000, 1250, 1500, 1750, 2000 \right].$}}\label{A5:fig4}
\end{figure}

\textcolor{Crimson}{\subsection{Example 3} \label{A5:Ex3}}

It is well--known that the singularity of the inverse of the autocovariance operator $C$  increases, when the rate of convergence to zero of  the eigenvalues of $C$ indicates a faster decay velocity,  as in this example.    Specifically, here,

\begin{eqnarray}
C_k  &=& \frac{1}{k^{2}},\quad k\geq 1. \nonumber 
\end{eqnarray}

As before, $\left\lbrace \rho_{k}^{2}, \ k \geq 1 \right\rbrace$ and $\left\lbrace \sigma_{k}^{2}, \ k \geq 1 \right\rbrace$ are generated from (\ref{A5:53})-(\ref{A5:55}). The truncation order $k_T$ satisfies
\begin{equation}
k_T = \lceil T^{1/\alpha} \rceil, \quad \displaystyle \lim_{T \to \infty} k_T = \infty, \quad \displaystyle \lim_{T \to \infty} \sqrt{T} C_{k_T} = \infty
\label{A5:trorder}
\end{equation}
  \noindent (see also the simulation study undertaken in  \cite{Alvarezetal17}, for the case of $\rho$ being a Hilbert--Schmidt operator). In particular,
  (\ref{A5:trorder}) holds for
  $\frac{1}{2} - \frac{2}{\alpha} > 0.$  Thus,  $\alpha > 4,$ and we consider $\alpha = 4.1,$ i.e., $k_T = \lceil T^{1/4.1} \rceil.$

\begin{figure}[H]
\centering
\includegraphics[width=0.8\textwidth]{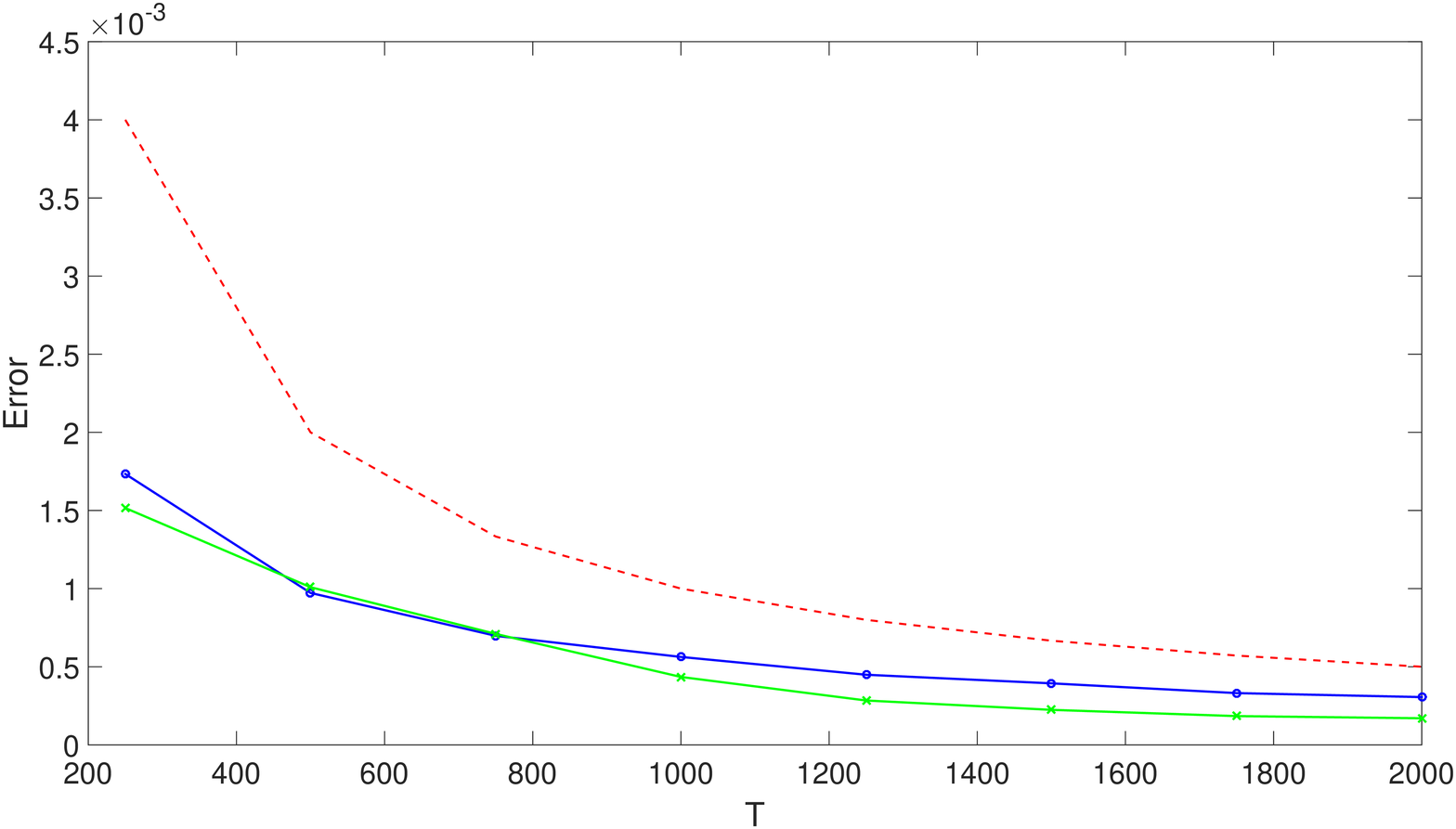}

\vspace{-0.25cm}
\caption[\hspace{0.7cm} Empirical quadratic functional errors for Example 3.]{\small{\textcolor{Crimson}{Example 3.}   Empirical functional mean--square estimation errors of
classical (blue circle line), and  Bayes (green cross line) componentwise ARH(1)
parameters estimators,  with $k_T = \lceil T^{1/\alpha} \rceil$,  $\alpha = 4.1$, for $N=1000$ replications of the ARH(1) values,
 against the curve $1/T$ (red dot line), for  $T= \left[ 250,  500, 750, 1000, 1250, 1500, 1750, 2000 \right]$.}}\label{A5:fig5}
\end{figure}

\begin{figure}[H]
\centering
\includegraphics[width=0.8\textwidth]{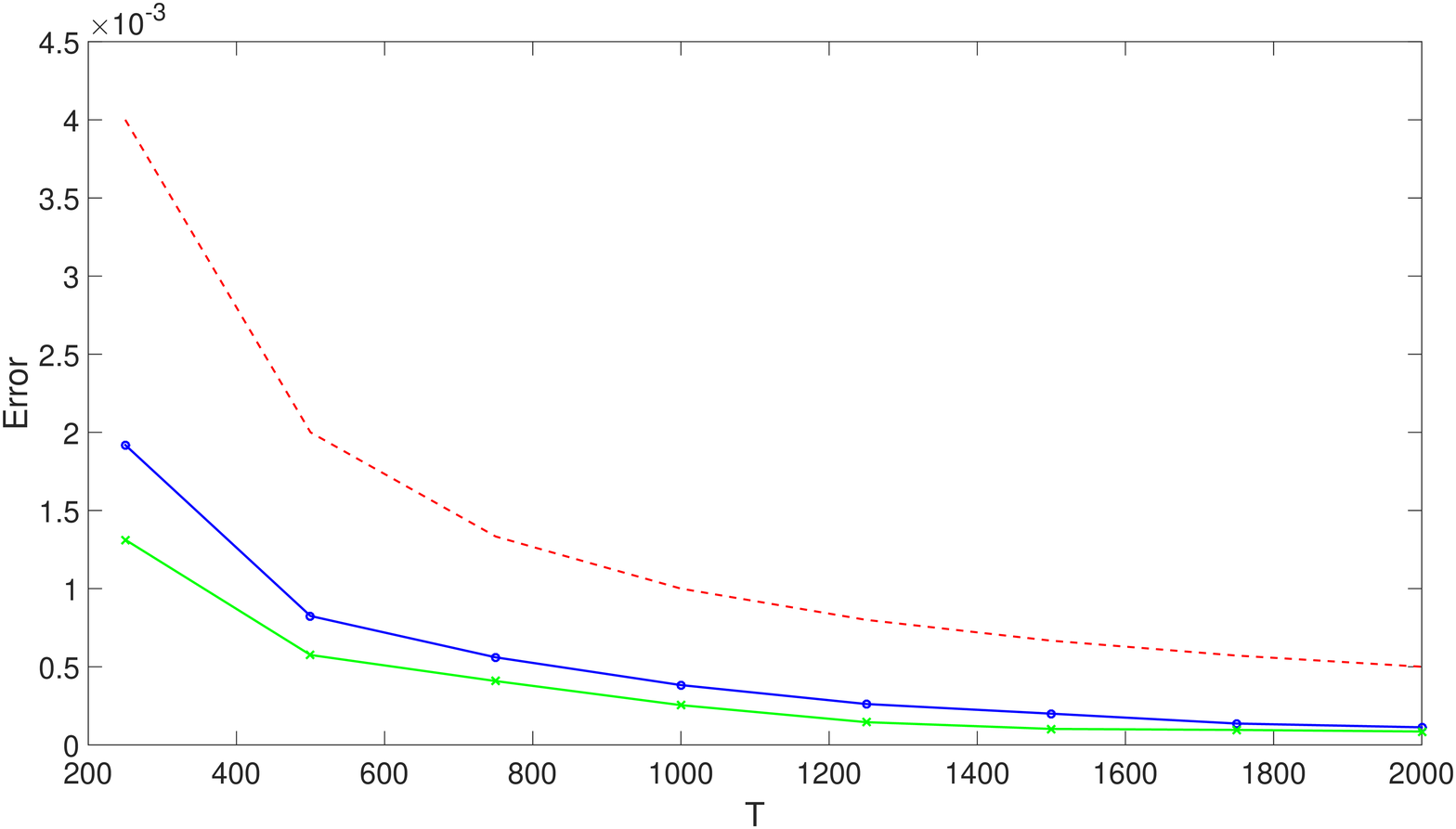}

\vspace{-0.25cm}
\caption[\hspace{0.7cm} Empirical quadratic functional errors for Example 3.]{\small{\textcolor{Crimson}{Example 3.}  Empirical functional mean--square prediction errors of
classical (blue circle line), and  Bayes (green cross line) componentwise ARH(1)
plug-in predictors,  with $k_T = \lceil T^{1/\alpha} \rceil$,  $\alpha = 4.1$, for $N=1000$ replications of the ARH(1) values,
against the curve $1/T$ (red dot line), for  $T= \left[ 250,  500, 750, 1000, 1250, 1500, 1750, 2000 \right]$.}}\label{A5:fig6}
\end{figure}

Tables \ref{A5:tab7}--\ref{A5:tab8}  show  the empirical
functional mean--square errors associated with $\widehat{\rho}_{T}$ and
$\widetilde{\rho}^{-}_{T},$   and with the corresponding
 ARH(1) plug--in predictors, respectively. As before, Figures \ref{A5:fig5}--\ref{A5:fig6} provide the
graphical representations, and  the values of the curve $1/T,$ for $T$ in (\ref{A5:ss}), with the aim  of illustrating the rate of convergence to zero of the truncated empirical functional mean quadratic errors.

\begin{table}[H]
\caption[\hspace{0.7cm} Empirical quadratic functional errors for \textcolor{Crimson}{Example 3} (with a sample-size dependent truncation parameter).]{\small{\textcolor{Crimson}{Example 3.} Empirical functional mean-square errors $EFMSE_{\overline{\rho}_{T}}$}.}
\label{A5:tab7}
\centering
\begin{small}
\begin{tabular}{|c|c|c|c|}
\hline Sample size  & $k_T$ & Classical estimator $\widehat{\rho}_{T} $ & Bayes estimator $\widetilde{\rho}^{-}_{T}$\\
\hline 
$250 $& $3$ & $1.73 \left(10 \right)^{-3}$ & $1.52\left(10 \right)^{-3}$\\
\hline
 $500 $& $4$ & $9.72 \left(10 \right)^{-4}$ &  $1.01 \left(10 \right)^{-3}$\\
 \hline
 $750 $& $5$ & $6.98 \left(10 \right)^{-4}$&  $7.10 \left(10 \right)^{-4}$\\
 \hline
 $1000 $& $5$ & $5.63 \left(10 \right)^{-4}$&  $4.35 \left(10 \right)^{-4}$\\
 \hline
 $1250 $& $5$ & $4.49 \left(10 \right)^{-4}$ &  $2.84 \left(10 \right)^{-4}$\\
 \hline
 $1500 $& $5$ & $3.94 \left(10 \right)^{-4}$&  $2.24 \left(10 \right)^{-4}$\\
 \hline
 $1750$ & $6$ & $3.31 \left(10 \right)^{-4}$ & $1.84 \left(10 \right)^{-4}$\\
 \hline
 $2000$ & $7$ & $3.05 \left(10 \right)^{-4}$ & $1.70 \left(10 \right)^{-4}$ \\
 \hline
\end{tabular}
\end{small}
\end{table}

\begin{table}[H]
\centering
\caption[\hspace{0.7cm} Empirical quadratic functional errors for \textcolor{Crimson}{Example 3} (with a sample--size dependent truncation parameter).]{\small{\textcolor{Crimson}{Example 3.} Empirical functional mean--square errors $EFMSE_{\overline{\rho}_{T} \left(X_T \right)}$}.}
\label{A5:tab8}

\begin{small}
\begin{tabular}{|c|c|c|c|}
\hline Sample size  & $k_T$ & Classical  predictor $\widehat{\rho}_{T} \left(X_T \right)$ & Bayes
predictor $\widetilde{\rho}^{-}_{T}\left(X_T \right)$\\
\hline 
$250$ & $3$ & $1.92 \left(10 \right)^{-3}$ &  $1.31 \left(10 \right)^{-3}$\\
\hline
 $500$ & $4$ & $8.24 \left(10 \right)^{-4}$&  $5.75 \left(10 \right)^{-4}$\\
 \hline
 $750$ & $5 $& $5.60 \left(10 \right)^{-4}$ &  $4.08 \left(10 \right)^{-4}$\\
 \hline
 $1000$ & $5 $& $3.52 \left(10 \right)^{-4}$ &  $2.54 \left(10 \right)^{-4}$\\
 \hline
 $1250$ & $ 5 $& $2.62 \left(10 \right)^{-4}$ &  $1.45 \left(10 \right)^{-4}$\\
 \hline
 $1500$ & $5$ &$ 2.00 \left(10 \right)^{-4}$ &  $1.02\left(10 \right)^{-4}$\\
 \hline
 $1750$ & $6 $& $ 1.37 \left(10 \right)^{-4}$ & $ 9.57\left(10 \right)^{-5}$\\
 \hline
 $2000$ &$ 6 $&$ 1.13 \left(10 \right)^{-4}$&  $8.55 \left(10 \right)^{-5}$\\
 \hline
\end{tabular}
\end{small}
\end{table}

\bigskip

In \textcolor{Crimson}{Examples 1--2} in \textcolor{Crimson}{Sections} \ref{A5:Ex1}--\ref{A5:Ex2}, where a common  fixed truncation order is considered, we can observe that  the  biggest values of the
empirical functional mean--square errors are located at  the smallest sample sizes, for which the number $k_{T}=5$ of parameters to be estimated is too large,    with a slightly worse performance for those sample
sizes, in \textcolor{Crimson}{Example 3} in \textcolor{Crimson}{Seciton} \ref{A5:Ex2}, where  a
slower decay velocity, than in  \textcolor{Crimson}{Example 1}, of the eigenvalues of the
autocovariance operator $C$ is considered. Note that, on the other hand, when a slower decay velocity of the eigenvalues of $C$ is given, a larger truncation order is required to explain a given percentage  of the functional variance.
 For the fastest rate of convergence to zero of the eigenvalues of the autocovariance operator $C,$ in  \textcolor{Crimson}{Example 3}, to compensate the singularity of the inverse covariance operator $C^{-1},$  a suitable truncation order $k_{T}$ is fitted, depending on the sample size $T,$ obtaining a  slightly better performance than in the previous cases, where a fixed truncation order is studied.

\textcolor{Crimson}{\section{Final comments}}

This paper addresses the case where the eigenvectors of $C$
 are known, in relation to the asymptotic efficiency and equivalence of $\widehat{\rho}_{j,T}$ and $\widetilde{\rho}_{j,T}^{-},$ and the associated plug-in predictors.  However, as shown in the simulation study undertaken in \cite{Alvarezetal17}, a similar performance is obtained in the case where the eigenvectors of $C$ are unknown (see also \cite{Bosq00} in relation to the asymptotic properties of the empirical eigenvectors of $C$).

 In  the cited references in the ARH(1) framework, the autocorrelation
operator is usually assumed to  belong to the Hilbert--Schmidt class. Here,
in the  absence of the compactness assumption (in particular, of the
Hilbert--Schmidt assumption) on the autocorrelation operator $\rho,$ singular autocorrelation kernels can be considered. As commented in the \textcolor{Crimson}{Section} \ref{A5:sec:1},
the singularity of $\rho$ is compensated by the regularity of the autocovariance kernel of the innovation process, as reflected in \textcolor{Aquamarine}{\textbf{Assumption A2B}}.

\textcolor{Crimson}{Theorem} \ref{A5:mr} establishes sufficient conditions for the asymptotic efficiency and equivalence of the proposed classical and Bayes diagonal componentwise parameter estimators of
$\rho,$ as well as of the associated  ARH(1) plug-in predictors (see \textcolor{Crimson}{Theorem} \ref{A5:mr2}).
The
simulation study illustrates the fact that the truncation order $k_{T}$ should be selected according to
 the rate of convergence to zero of the eigenvalues of the autocovariance operator, and depending on
 the sample size $T.$ Although, a fixed  truncation order, independently of $T,$ has also been tested in \textcolor{Crimson}{Examples 1--2}, where a compromise between the
 rate of convergence to zero of the eigenvalues, and the rate of increasing of the sample sizes is found.

\textcolor{Crimson}{\section{Supplementary Material: Bayesian estimation of real--valued autoregressive processes of order one}
\label{A5:Supp1}}

 In this section, we consider the Beta--prior--based
Bayesian estimation of the autocorrelation coefficient $\rho$ in a standard AR(1) process. Namely, the generalized maximum likelihood estimator of such a parameter is computed, when a beta prior is assumed for $\rho.$ In the ARH(1) framework,
 we have adopted this  estimation procedure in the approximation of the diagonal coefficients $\{ \rho_{k},\ k\geq 1\}$ of operator $\rho$ with respect to
 $\{\phi_{k}\otimes \phi_{k},\ k\geq 1\},$ in a Bayesian componentwise context.  Note that   we also denote by $\rho$ the autocorrelation coefficient of an AR(1) process, since there is no place for confusion here.

 Let $\{X_{n},\ n\in \mathbb{Z}\}$ be an AR(1) process satisfying \begin{equation}X_{n}=\rho
X_{n-1}+\varepsilon_{n},\quad n\in \mathbb{Z}, \nonumber 
\end{equation}
\noindent where $0<\rho<1,$ and $\{\varepsilon_{n},\ n \in \mathbb{Z}\}$ is a real--valued Gaussian white
noise; i.e.,  $\varepsilon_{n}\sim \mathcal{N}(0,\sigma^{2}),$ $n\in \mathbb{Z},$ are independent Gaussian random
variables, with $\sigma
>0.$   Here, we will use the conditional likelihood, and assume that $(x_{1},\dots,x_{n})$ are observed
for $n$ sufficiently large to ensure that the effect of the random initial condition  is negligible. A beta
distribution with shape parameters $a>0$ and $b>0$ is considered as
 a-priori distribution on $\rho,$  i.e., $\rho \sim \mathcal{B} (a,b).$
Hence, the distribution of $(x_{1},\dots,x_{n},\rho)$ has density

$$\widetilde{L}=\frac{1}{(\sigma \sqrt{2\pi})^{n}}\exp\left(-\frac{1}{2\sigma^{2}}\sum_{i=1}^{n}(x_{i}-\rho x_{i-1})^{2}\right)\rho^{a-1}(1-\rho )^{b-1}
\frac{\boldsymbol{1}_{\{0<\rho<1\}}}{\mathbb{B}(a,b)},$$ \noindent where
$$\mathbb{B}(a,b)=\frac{\Gamma(a)\Gamma(b)}{\Gamma (a+b)}$$\noindent
is the beta function.

We first compute the solution to the equation
\begin{eqnarray}
0=\frac{\partial \ln \widetilde{L}}{\partial \rho
}&=&\frac{\partial}{\partial \rho
}\left[-\frac{1}{2\sigma^{2}}\sum_{i=1}^{n}(x_{i}-\rho x_{i-1})^{2}
+(a-1)\ln \rho+(b-1)\ln (1-\rho)\right]\nonumber\\
&=& -\frac{1}{2\sigma^{2}}\sum_{i=1}^{n}\left( -2x_{i-1}(x_{i}-\rho
x_{i-1})\right)+\frac{a-1}{\rho}-\frac{b-1}{1-\rho}\nonumber\\
&=& \frac{\alpha_{n} }{\sigma^{2}}-\frac{\rho}{\sigma^{2}}\beta_{n}
+\frac{a-1}{\rho}-\frac{b-1}{1-\rho}, \nonumber 
\end{eqnarray}
\noindent where $$\alpha_{n} =\sum_{i=1}^{n}x_{i-1}x_{i}, \quad \beta_{n}
=\sum_{i=1}^{n}x_{i-1}^{2}.$$
 Thus, the following equation must be
solved:
\begin{eqnarray}
0 &=&\frac{\rho (1-\rho)\alpha_{n}
}{\sigma^{2}}-\frac{\rho^{2}(1-\rho)}{\sigma^{2}}\beta_{n} +
(a-1)(1-\rho)-\rho (b-1)\nonumber\\
0 &=&\frac{\beta_{n}}{\sigma^{2}}\rho^{3}-\frac{\alpha_{n}
+\beta_{n}}{\sigma^{2}}\rho^{2}+\left(\frac{\alpha_{n}
}{\sigma^{2}}-[a+b]+2\right)\rho+(a-1).\nonumber 
\end{eqnarray}

\bigskip

\begin{itemize}
\item[\textcolor{Crimson}{Case 1}] Considering $a=b=1,$ and $\sigma^{2}=1,$ we obtain
the solution
$$\widetilde{\rho}_{n}=\frac{\displaystyle \sum_{i=1}^{n}x_{i-1}x_{i}}{\displaystyle \sum_{i=1}^{n}x_{i-1}^{2}}.$$
\item[\textcolor{Crimson}{Case 2}] The general case where $b>1$ is more intricate, since the
solutions are $\widetilde{\rho}_{n}=0,$ and
\begin{eqnarray}\widetilde{\rho }_{n}&=&\frac{1}{2\beta_{n}}\left[ (\alpha_{n}
+\beta _{n})\pm \sqrt{(\alpha_{n} -\beta_{n} )^{2}-4\beta_{n}
\sigma^{2}[2-(a+b)]}\right]\nonumber\\
&=& \frac{ \displaystyle \sum_{i=1}^{n}x_{i-1}x_{i}+x_{i-1}^{2}}{
2\displaystyle \sum_{i=1}^{n}x_{i-1}^{2}} \nonumber \\
&\pm &\frac{
\sqrt{\left[\displaystyle \sum_{i=1}^{n}x_{i-1}x_{i}-x_{i-1}^{2}\right]^{2}-4\sigma^{2}\left[\displaystyle \sum_{i=1}^{n}x_{i-1}^{2}\right][2-(a+b)]}}{
2 \displaystyle \sum_{i=1}^{n}x_{i-1}^{2}}. \nonumber
\end{eqnarray}

\item[\textcolor{Crimson}{Case 3}] For $\sigma^{2}=a=1,$  we have
\begin{eqnarray}\widetilde{\rho }_{n}&=&\frac{1}{2\beta_{n} }\left[
(\alpha_{n} +\beta_{n} )\pm \sqrt{(\alpha_{n} -\beta_{n} )^{2}-4\beta_{n} (1-b)}\right]\nonumber\\
&=&\frac{1}{2\displaystyle \sum_{i=1}^{n}x_{i-1}^{2}}\left[\sum_{i=1}^{n}x_{i-1}x_{i}+x_{i-1}^{2}\right] \nonumber \\
& \pm & \sqrt{\left[\displaystyle \sum_{i=1}^{n}x_{i-1}x_{i}-x_{i-1}^{2}\right]^{2}-4\left[\displaystyle \sum_{i=1}^{n}x_{i-1}^{2}\right](1-b)}. \nonumber 
\end{eqnarray}
\end{itemize}

\textcolor{Crimson}{\section{Supplementary Material 2: strong--ergodic AR(1) processes}
\label{A5:Supp2}}

This section collects some strong--ergodicity results applied in this paper, for real--valued weak--dependent  random sequences. In particular, their application to the  AR(1) case is considered.

A real--valued stationary process $\left\lbrace Y_n, \ n \in \mathbb{Z} \right\rbrace$ is strongly--ergodic (or ergodic in an almost surely sense), with respect to ${\rm E} \left\lbrace f \left(Y_0,\ldots,Y_{n-1} \right) \right\rbrace$ if, as $n\rightarrow \infty,$
\begin{equation}
\frac{1}{n-k} \displaystyle \sum_{i=0}^{n-1-k} f \left( Y_i, \ldots, Y_{i+k} \right) \longrightarrow^{a.s.} {\rm E} \left\lbrace f \left(Y_0,\ldots,Y_{n-1} \right) \right\rbrace, \quad k \geq 0. \nonumber
\end{equation}

In particular, the following lemma provides sufficient condition to get the strong--ergodicity for all second--order moments (see, for example,  \cite[Theorem 3.5.8]{Stout74} and \cite[p. 495]{Billingsley95}).

\bigskip
\begin{lemma}
\label{A5:lemma5}
\textit{Let $\left\lbrace \widetilde{\varepsilon}_n, \ n \in \mathbb{Z} \right\rbrace$ be an i.i.d. sequence of real--valued random variables. If $f:~\mathbb{R}^{\infty} \longrightarrow \mathbb{R}$ is a measurable function, then
\begin{equation}
Y_n = f \left( \widetilde{\varepsilon}_n, \widetilde{\varepsilon}_{n-1}, \ldots \right), \quad n \in \mathbb{Z}, \nonumber
\end{equation}
is a stationary and  strongly--ergodic process for all second--order moments.}
\end{lemma}

\bigskip

\textcolor{Crimson}{Lemma} \ref{A5:lemma5} is now applied to the invertible  AR(1) case, when the innovation process is white noise.

\bigskip

\begin{remark}
\label{A5:remark2}
\textit{If $\left\lbrace Y_n, \ n \in \mathbb{Z} \right\rbrace$ is a  real--valued zero--mean  stationary AR(1) process
\begin{equation}
Y_n = \rho Y_{n-1} + \widetilde{\varepsilon}_{n},\quad \rho \in \mathbb{R}, \quad \left| \rho \right| < 1, \quad n \in \mathbb{Z}, \nonumber
\end{equation}
\noindent where $\left\lbrace \widetilde{\varepsilon}_n, \ n \in \mathbb{Z} \right\rbrace$ is  strong white noise, we can define the measurable (even continuous) function
\begin{equation}
f \left( a_0, a_1, \ldots \right) = \displaystyle \sum_{k=0}^{\infty} \rho^{k} a_k, \nonumber
\end{equation}
\noindent such that, from \textcolor{Crimson}{Lemma} \ref{A5:lemma5} and for each $n \in \mathbb{Z}$,
\begin{equation}
Y_n = \displaystyle \sum_{k=0}^{\infty} \rho^{k} \widetilde{\varepsilon}_{n-k} = f \left( \widetilde{\varepsilon}_{n}, \widetilde{\varepsilon}_{n-1}, \ldots \right), \nonumber
\end{equation}
\noindent is a stationary and strongly--ergodic process for all second--order moments.}
\end{remark}

\bigskip

In the results derived in this paper, \textcolor{Crimson}{Remark} \ref{A5:remark2} is applied, for each $j\geq 1,$  to the real--valued zero--mean  stationary AR(1) processes $$\left\lbrace X_{n,j}=\left\langle X_{n},\phi_{j}\right\rangle_{H}, \ n \in \mathbb{Z} \right\rbrace,$$ with $\{X_{n} ,\ n\in \mathbb{Z}\}$ now representing an ARH(1) process.

\bigskip

\begin{corollary}
\label{A5:corollary1}
\textit{Under \textcolor{Aquamarine}{\textbf{Assumptions A1--A2}}, for each $j\geq 1,$ let us consider the real--valued zero--mean  stationary AR(1)  process $\left\lbrace X_{n,j}=\left\langle X_{n},\phi_{j}\right\rangle_{H}, \ n \in \mathbb{Z} \right\rbrace$, such that, for each $n \in \mathbb{Z}$
\begin{equation}
X_{n,j} = \rho_j X_{n-1,j} + \varepsilon_{n,j}, \quad \rho_j \in \mathbb{R}, \quad \left| \rho_j \right| < 1, \nonumber
\end{equation}
\noindent  Here,  $\left\lbrace \varepsilon_{n,j}, \ n \in \mathbb{Z} \right\rbrace$ is a real-valued strong white noise, for any $j \geq 1$. Thus, for each $j \geq 1$, $\left\lbrace X_{n,j}, \ n \in \mathbb{Z} \right\rbrace$ is a stationary and strongly-ergodic process for all second-order moments. In particular, for any $j \geq 1$, as $n\rightarrow \infty,$
\begin{eqnarray}
\widehat{C}_{n,j} = \frac{1}{n} \displaystyle \sum_{i=1}^{n} X_{i-1,j}^{2} &\longrightarrow^{a.s.} &C_j = {\rm E} \left\lbrace X_{i-1,j}^{2} \right\rbrace,\ i\geq 1, \nonumber \\
\widehat{D}_{n,j} = \frac{1}{n-1} \displaystyle \sum_{i=1}^{n} X_{i-1,j}X_{i,j} & \longrightarrow^{a.s.} &D_j = {\rm E} \left\lbrace X_{i-1,j} X_{i,j} \right\rbrace, \ i\geq 1.\nonumber 
\end{eqnarray}}
\end{corollary}

\textcolor{Crimson}{\section*{\textbf{Acknowledgments}}}

\textcolor{Aquamarine}{\textbf{This work has been supported in part by projects MTM2012--32674 and MTM2015--71839--P (co-funded by Feder funds), of the DGI, MINECO, Spain.}}

\vspace{0.5cm}
\renewcommand\bibname{\textcolor{Crimson}{\textit{\textbf{References}}}}

\bibliographystyle{dinat}
\bibliography{Biblio}

\end{document}